\input amstex
\documentstyle{amsppt}

\comment
\voffset-3pc
\magnification=\magstep1
\baselineskip=24true pt
\endcomment

\topmatter
\title A rigorous real time Feynman Path Integral and Propagator
       \endtitle
\author Ken Loo \endauthor
\address PO Box 9160, Portland, OR.   97207
         \endaddress
\email look\@sdf.lonestar.org \endemail
\thanks The author would like dedicate this paper to 
        Karen Soohoo and Photine Tsoukalas; give thanks
        to the referees for their constructive
        comments; and give thanks to Ellen Hambrick
        for her help on references.
        \endthanks
\abstract We will derive a rigorous real time propagator
   for the Non-relativistic 
   Quantum Mechanic $L^2$ transition probability amplitude and
   for the Non-relativistic wave function.  
   The propagator will be explicitly given in terms of
   the time evolution operator.  
   The derivation will be for all self-adjoint nonvector
   potential Hamiltonians.  For systems with potential that
   carries at most a finite number of singularity and discontinuities,
   we will show that our propagator can be written in the form of
   a rigorous real time, time sliced Feynman path integral 
   via improper Riemann integrals.  We will also derive 
   the Feynman path integral in Nonstandard Analysis Formulation. 
   Finally, we will compute the propagator for the harmonic oscillator
   using the Nonstandard Analysis Feynman path 
   integral formuluation; we will compute the
   propagator without using any knowledge of classical properties of
   the harmonic oscillator.
\endabstract
\leftheadtext{A Rigorous Real Time Feynman Path Integral and Propagator}
\rightheadtext{A Rigorous Real Time Feynman Path Integral and Propagator}
\endtopmatter
\document

\define\fac#1#2{\left(\frac {m}{2\pi{}i\hbar\epsilon}\right)^
    {\frac{#1}2\left(#2 + 1 \right)}}

\define\summ#1#2#3#4{\frac {i\epsilon}{\hbar}\sum\limits_{j=1}
    ^{{#1}+1}\left[\frac {m}{2}\left(\frac {{#2} - {#3}}
    {\epsilon}\right)^2\! #4\right]}

\define\pd#1#2{\dfrac{\partial#1}{\partial#2}}
\define\spd#1#2{\tfrac{\partial#1}{\partial#2}}
\define\ham1#1#2{\dfrac{-\hbar^2}{2m}\Delta_{#1} #2}

\subhead
{\bf I. Introduction}\endsubhead
Since Feynman's invention of the path integral, much
research have been done to make the real time Feynman
path integral mathematically rigorous(see [6], [9], [10],
[13],[18], [19], and [20]).  In physics, the real time, time sliced
Feynman path integral is formally given by
(see [3],[4], and [5])
$$\align
  {}&\bar K_t\left(\vec x, \vec x_0\right) = \lim_{k\to\infty}
  w_{n,k}\int_{r\Bbb R^{\left(k-1\right)n}}
  \text{exp}\left[\dfrac{i\epsilon}{\hbar}
   S_k\left(\vec x=\vec x_k,...,\vec x_o\right)\right]
   d\vec x_1...d\vec x_{k-1},\tag1.1\\
  {}&\vec x_0=\vec q_0, \quad \vec x_{k+1}=\vec q, \quad
     \epsilon =\dfrac{t}{k}, \quad
   w_{n,k} = \left(\dfrac{m}{2i\pi\hbar\epsilon}\right)^
   \frac{n(k+1)}{2}, \\
   {}&S\left\{\vec x_{k+1}\dots\vec x_{0}\right\} =
   \sum_{j=1}^{k+1}\left[\dfrac{m}{2}
   \left(\dfrac{\vec x_{j} - \vec x_{j-1}}{\epsilon}\right)^2 -
   V\left(\vec x_j\right)
   \right], \\
  {}&\left[\text{exp}
     \left(\dfrac{-it\bar H}{\hbar}\right)
  \psi\right]\left(\vec x\right) =
  \int \bar K_t\left(\vec x, \vec x_0\right)
  \psi\left(\vec x_0\right) d\vec x_0 , \\
  {}&\left[\text{exp}
  \left(\dfrac{-it\bar H}{\hbar}\right)
  \delta\left(\vec y - \vec x_0\right)\right]\left(\vec x\right) =
  \bar K_t\left(\vec x, \vec x_0\right),
\endalign
$$
where the integral of the first equation
in (1.1) is an improper Riemann integral,
and the last line in (1.1) is the evolution
operator operating on the dirac delta function's
$\vec y$ variable.
It is well known that mathematical rigor of (1.1) is lacking,
and we know that an integral over path space in real
time can not be well defined with measure theory(see [6]).

The problems with the objects in equation 1.1
are that we do not know if the
improper Riemann integrals exist, we do not know if the
$k$ limit exists, and we do not know if the Feynman
path integral in 1.1 produces the
propagator.
In his paper(see [11] footnote 13), Feynman observed that
by using wave functions, ill-defined oscillatory integrals
can be given rigorous meaning.  With this observation,
we will reformulate equation 1.1 into rigorous mathematical
objects that represent the propagator.

From mathematics,
we know that for some values of $t$, some propagators must
be treated as distributions; the harmonic oscillator is
one example(see [6] and [7]).
Also, $\bar K_t\left(\vec x,\vec x_0\right)$ given in (1.1)
is a function of $\vec x$ and $\vec x_0$.  Thus, it is
natural to consider $\bar K_t$ as a tempered distribution
on the class of Schwartz test functions
$S\left(\Bbb R^n\times\Bbb R^n\right)$.  The space of
square integrable functions is a subset of the space of
tempered distributions.  If we consider the wave function
as a distribution and take its inner produce with a test
function, we can formally use (1.1) and get

$$\align
  {}&\int_{\Bbb R^n} \phi\left(\vec x\right)\left[
     \text{exp}\left(\dfrac{-it\bar H}{\hbar}\right)\psi\right]
     \left(\vec x\right)
     d\vec x\,\, = \tag1.2 \\
  {}&\int_{\Bbb R^n} \phi\left(\vec x\right)
  \int \bar K_t\left(\vec x, \vec x_0\right)
  \psi\left(\vec x_0\right) d\vec x_0 d\vec x = \\
  {}&\int \bar K_t\left(\vec x, \vec x_0\right)
     \phi\left(\vec x\right)\psi\left(\vec x_0\right)
     d\vec x d\vec x_0.
\endalign
$$
Equation 1.2 will be the fifth theorem in section II.

As for the formal evolution of the delta function in (1.1),
let us formally consider the following equation.
$$\align
   {}&\lim_{\eta ,\gamma\to 0}
      K\left(\vec x, \vec x_{0},\eta ,\gamma ,t\right) =
      \tag1.3\\
   {}&\lim_{\eta ,\gamma\to 0}
      \int_{\Bbb R^n} G_{x}\left(\vec y, \eta\right)\left[
      \text{exp}
      \left(\dfrac{-it\bar H}{\hbar}\right)
      F_{\vec x_0}\left(\vec z, \gamma\right)
      \right]\left(\vec y\right)
      d\vec y   = \\
  {}&\int\delta\left(\vec x - \vec y\right)
   \left[\text{exp}
  \left(\dfrac{-it\bar H}{\hbar}\right)
  \delta\left(\vec z - \vec x_0\right)\right]\left(\vec y\right)
  d\vec y = \\
  {}&\left[\text{exp}
  \left(\dfrac{-it\bar H}{\hbar}\right)
  \delta\left(\vec z - \vec x_0\right)\right]\left(\vec x\right),
\endalign
$$
where the functions $K, F$ and $G$ are given by equation. 2.1. 
If we are going to take the propagator as a distribution in
the sense of (1.2), we might consider the limit in (1.3) as a distribution
limit.  Doing so produces the second theorem and in some sense
the fourth theorem in section II(equations 2.2, 2.3, 2.6a, and
2.6b).

In mathematics, there exists a rigorous formulation for a
real time, time sliced Feynman path
integral(see [7] and [8]), it reads
$$\align
  {}&\left[
  \text{exp}
  \left(\dfrac{-it\bar H}{\hbar}\right)\psi\right]\left(\vec x\right)
   = \tag1.4 \\
   {}&\lim_{k\to\infty}
     w_{n,k}
   \int_{\Bbb R^{kn}}
   \text{exp}\left[\dfrac{i\epsilon}{\hbar}
   S_k\left(\vec x_k = \vec x,\dots ,\vec x_0\right)
   \right]\psi\left(\vec x_0\right)
   \,d\vec x_0\dots d\vec x_{k-1},
\endalign
$$
where $\psi\in L^2$, the integral in (1.4) is an improper
Lebesgue integral with convergence taken in the $L^2$
topology, and the $k$ limit in (1.4) is taken in the
$L^2$ topology.  Comparing (1.4) and (1.1), we see that
rigorously we have a Feynman path integral that has all
convergence taken in $L^2$ topology while in physics, formal
improper Riemann integral and pointwise convergence is favored.

What we will do is convert all convergences in $L^2$ topology
into pointwise convergences in $t$.
The idea is to use a wave function as a convergence factor
as observed by Feynman.  For simplicity,
suppose $f\left(x\right)\in L^2\left(\Bbb R\right),
g\left(x, y\right)
\in L^2\left(\Bbb R\times\Bbb R\right)$ are such that
they are bounded and continuous.  Further, suppose that
both
$$\align
  {}&h\left(x\right) = \int_{-a}^{b}
   g\left(x, y\right)dy,  \tag1.5\\
  {}&p\left(x\right) =
   \lim_{a,b\to\infty}\int_{-a}^{b}
   g\left(x, y\right)dy,
\endalign
$$
are in $L^2\left(\Bbb R\right)$ as a function of
$x$.  In (1.5), we take the integrals to
be Lebesgue integrals and the limits are taken
independent of each other in the $L^2$ norm.
Notice that for $p\left(x\right)$, we can interpret the
integral
as an improper Lebesgue integral with convergence
in the $L^2$ topology.  Let us denote $\chi_{[-c,d]}$
to be the characteristic function on $[-c, d]$.
Schwarz's inequality then implies
$$\align
  {}&\Bigg|\int_{\Bbb R} f\left(x\right)p\left(x\right)dx -
  \int_{-c}^{d}
  \int_{-a}^{b}
  f\left(x\right)g\left(x, y\right)dx dy\Bigg| \leq \tag1.6\\
  {}&||f||_2\,||p - h||_2 + ||f - \chi_{[-c,d]}f||_2\,||h||_2
    \to 0.
\endalign
$$

Thus, we can write
$$
  \int_{\Bbb R} f\left(x\right)p\left(x\right)dx =
  \lim_{a,b,c,d\to\infty}\int_{-c}^{d}
  \int_{-a}^{b}f\left(x\right)g\left(x, y\right)dx dy, \tag1.7
$$
where the limits are all taken independent of each other.
Since $f$ and $g$ are bounded and continuous,
the Lebesgue integral over $[-a, b]\times [-c, d]$
in (1.7) can be replaced by
a Riemann integral.  Since the limits are taken independent
of each other, we can then interpret the right hand-side
of (1.7) as an improper Riemann integral.  If $f$ and
$g$ carry singularities and discontinuities,
care must be taken in the
region of integration so that
the replacement of Lebesgue integral with Riemann integrals
can be done.

The technique of converting $L^2$ limits into pointwise
limits as illustrated above is what we will use to prove all the theorems
in the next section.
It is the foundation of this work.

\subhead{\bf II. Results}\endsubhead
In his paper(see [11] footnote 13), Feynman observed that 
by using wave functions, ill-defined oscillatory integrals
can be given rigorous meaning.  With this observation,
we will reformulate equation 1.1 into a rigorous mathematical 
object that represents the propagator.

The goal of this paper is the following. 
First, we will elaborate on
Feynman's observation and use wave functions to provide
a convergence factor in the derivation of a real time propagator that
takes the form of an $L^2$ transition probability 
amplitude.  We will use wave functions to derive 
a real time, time sliced Feynman path integral.  We will
derive two Nonstandard Analysis formulations of the
time sliced Feynman path integral.  
Finally, we will compute the propagator of the harmonic oscillator
using our Nonstandard Feynman path integral representation.
We will assume that the reader is familiar
with Nonstandard Analysis(see [13]-[17] and references within).  

The usual idea in using
Nonstandard Analysis is to replace the time slice limit by
a standard part(see [9],[13],[18],[19] and
references within).  We will derive a Nonstandard formulation
that transfers the time slice limit into the nonstandard
world and standard part is taken on infinitesimal parameters
in wave functions.  It was shown in
[19] that for the harmonic oscillator, equation 1.1 can be cast
into the language of Nonstandard Analysis where the time slice 
limit is replaced by a standard part.  Further, using Nonstandard
Analysis methods, one can rigorously
compute the harmonic oscillator propagator without having
prior knowledge of the classical path.  We will follow the 
approach of [19] in the computations of this paper.  
In [19], we do not know if
equation 1.1 is the propagator apriori; we are satisfied
because the computation produced the correct results.
In this paper, we will have a Feynman path integral representation
that is known to produce the propagator and we will use it
to compute the harmonic oscillator propagator.

What we will show is the following.
Let $H = \dfrac{-\hbar^2}{2m}\Delta + V\left(\vec x\right) =
H_0 + V\left(\vec x\right)$ be 
essentially self-adjoint and the domain of $H$ 
contains the Schwartz space of rapidly decreasing
test functions. Denote the closure
of $H$ by $\bar H$.  Let $t > 0$ and let
$$
\align
 {}&F_{\vec x}\left(\vec y, \gamma\right) =
    F\left(\vec x, \vec y, \gamma\right) =
    \left(\dfrac{m}{2\pi\hbar\gamma}\right)^
    \frac{n}{2}
    \text{ exp }\left[\dfrac{-m\gamma}{2\hbar}
  \left(\dfrac{\vec x - \vec y}{\gamma}\right)^2
  \right],\quad\gamma > 0\tag2.1\\
  {}&G_{\vec x}\left(\vec y,\eta\right) =
     G\left(\vec x, \vec y,\eta\right) =
   \left(\dfrac{m}{2\pi\hbar\eta}\right)^
    \frac{n}{2}
    \text{ exp }\left[\dfrac{-m\eta}{2\hbar}
  \left(\dfrac{\vec x - \vec y}{\eta}\right)^2
  \right], \quad\eta > 0 \\
 {}&K\left(\vec x, \vec x_{0},\eta ,\gamma ,t\right) = 
  \int_{\Bbb R^n} G_{\vec x}\left(\vec y, \eta\right)\left[
  \text{exp}
  \left(\dfrac{-it\bar H}{\hbar}\right)
   F_{\vec x_0}\left(\vec z, \gamma\right) 
   \right]\left(\vec y\right)
  d\vec y.  \\
\endalign
$$
The notation for $K\left(\vec x, \vec x_{0},\eta ,\gamma ,t\right)$ 
is that the evolution operator operates on the $\vec z$ variable
while leaving $\vec x_0$ fixed and the result is a function of 
$\vec x_0$ and $\vec y$; finally, the $\vec y$ variable is
integrated against $G$.  We point out that $K$ in (2.1) is in the form
of an $L^2$ transition probability amplitude, but neither $F$ 
nor $G$ are wave functions since they are not normalized to 1 in
the $L^2$ norm.  Also, the form of $K$ is similar to the form of the
propagator given in [12] where Prugovecki provides a theory of stochastic
Quantum Mechanics.  It will be interesting to see the relationship between
(2.1) and the stochastic propagator derived by Prugovecki.  One immediate 
difference between (2.1) and Progovecki's formulation is that (2.1) stays 
within the popular representation of Quantum Mechanics where as Prugovecki
uses a different representation(see [12]).

The existence of 
$K\left(\vec x, \vec x_{0},\eta ,\gamma ,t\right)$
is immediate since both functions in the
integrand are in $L^2$.
We will show that 
\proclaim{First Theorem} 
$K\left(\vec x, \vec x_{0},\eta ,\gamma ,t\right)$
is continuous as a function of $\left(\vec x,
\vec x_0\right) \in \Bbb R^{2n}$ and it is uniformly
bounded as a function of $\left(\vec x,
\vec x_0\right) \in \Bbb R^{2n}$.
\endproclaim

The kernel in the first
theorem will play the role of an integral kernel in the 
following sense.
\proclaim{Second Theorem Part a} Let $\phi ,\psi\in L^2\left(\Bbb R^n\right)$
Let $H$ be essentially self-adjoint, then
$$\align
  {}&\int_{\Bbb R^n} \phi\left(\vec x\right)\left[
  \text{exp}
  \left(\dfrac{-it\bar H}{\hbar}\right)\psi\right]\left(\vec x\right)
  d\vec x  = \tag2.2 \\
  {}&\lim_{\eta ,\gamma\to 0}
   \bar{\int}_{\Bbb R^{2n}}\phi\left(\vec x\right)
   \psi\left(\vec x_0\right)
   K\left(\vec x, \vec x_{0},\eta ,\gamma ,t\right)  
   \,d\vec x_0 d\vec x,
\endalign
$$
where the notation for the integral in the right hand side of
the equality in (2.2) means an improper Lebesgue integral with
the convergence at infinity taken pointwise in $t$(see
equation 5.10 for more details), and
the limits are taken independent of each other and pointwise in $t$.
\endproclaim
Notice that in the above theorem, the convergence of the improper Lebesgue
integral is pointwise in $t$ as opposed to convergence in the
$L^2$ topology in equation 1.4.  A pointwise convergence might
provide computational advanages.

We will not attempt to pass the limits in (2.2) inside the
integral since some real time propagators does not exist
as a function for all time and must be treated as distributions
(see [1]).  
We will make a connection between 
$K\left(\vec x, \vec x_{0},\eta ,\gamma ,t\right)$
and the theory of distributions(see remark 6.2 and equation 6.4).  
On the other hand, we would be like to be able to pass the limits inside
the improper Lebesgue integral when the kernel in the first theorem
and the propagator for the evolution are well behaved.
The next theorem provides us with that opportunity.

\proclaim{Second Theorem Part b} Let $\phi ,\psi\in 
L^2\bigcap
L^1$.  Let $H$ be essentially self-adjoint, 
then
$$\align
  {}&\int_{\Bbb R^n} \phi\left(\vec x\right)\left[
  \text{exp}
  \left(\dfrac{-it\bar H}{\hbar}\right)\psi\right]\left(\vec x\right)
  d\vec x  = \tag2.3 \\
  {}&\lim_{\eta ,\gamma\to 0}
   \int_{\Bbb R^{2n}}\phi\left(\vec x\right)
   \psi\left(\vec x_0\right)
   K\left(\vec x, \vec x_{0},\eta ,\gamma ,t\right)   
   \,d\vec x_0 d\vec x,
\endalign
$$
where the integral in the right hand side of the equality
is a Lebesgue integral and all limits are taken independent
of each other and pointwise in
$t$. 
\endproclaim
In the above theorem, 
the first theorem and the fact that the wave functions are
in $L^1$ provide us the opportunity to pass the limits inside
the integral and produce the propagator for the evolution.  
We will do this for the harmonic oscillator Hamiltonian.  
Further, for the purpose of passing the limits,
we will not attempt to generalize the wave functions to all
of $L^2$ since integrating the propagator against two arbitrary
$L^2$ wave functions in the sense of equation 1.2 is not always
well defined; the free evolution propagator is one such example. 

The kernel in (2.1) can  be explicitely represented
by a time sliced Feynman path integral.  
\proclaim{Third Theorem} 
Let $H$ be essentially self-adjoint,  
and the potential $V$ be such that 
it has at most a finite number of discontinuities
and singularities.  Let
$$\align
   {}&w_{n,k} = \left(\dfrac{m}{2i\pi\hbar\epsilon}\right)^
    \frac{nk}{2}, \epsilon = \dfrac{t}{k}, \tag2.4\\
   {}&S_k\left(\vec x_{k+1},\dots ,\vec x_1\right) =
   \sum\limits_{j=2}
   ^{k+1}\left[\dfrac {m}{2}\left(\dfrac {\vec x_j - \vec x_{j-1}}
   {\epsilon}\right)^2\! - V\left(\vec x_j\right)\right],
\endalign
$$
then
$$
\align
 {}&K\left(\vec x, \vec x_{0},\eta ,\gamma ,t\right) = \tag2.5\\
 {}&\lim_{k\to\infty}w_{n,k}
    \int_{r\Bbb R^{\left(k+1\right)n}}
    F_{\vec x_0}\left(\vec x_1, \gamma\right) 
     \text{exp}\left[\dfrac{i\epsilon}{\hbar}
    S_k\left(\vec x_{k+1},...,\vec x_1\right)\right]
    G_{\vec x}\left(\vec x_{k+1},\eta\right) 
   d\vec x_1...d\vec x_{k+1}. \\
\endalign
$$
In (2.5), the integral is an improper Riemann integral and
the $k$ limit is taken pointwise in $t$.  
\endproclaim
In the third theorem, there is no restriction on the type
of discontinuities and singularities on the potential as
long as the Hamiltonian is essentially self-adjoint and
by improper Riemann integral, we mean a Riemann integral
with convergence at infinity taken pointwise in $t$.
Further, it is not necessary to put the restriction on potential
in the above theorem and work with improper Riemann integrals
if one is willing to work with improper Lebesgue integral as
in part a of the second theorem(see remark 3.2), but for our purpose of
computing the harmonic oscillator propagator(see section VIII),
we will formulate the third theorem as above.

The problem with formulating a real time,
time sliced Feynman path integrals is that 
Fubini's theorem can not be applied due to the 
oscillatory nature of the integrand. 
We will see that the application of Fubini's
theorem can be justified in the derivation of (2.5) because the
functions $F$ and $G$ play the role of convergence factors as
Feynman pointed out.

The propagator is usually formulated for the wave function.
If we wish to work with the wave function, we have the
following.
\proclaim{Fourth Theorem Part a} Let 
$\psi\in L^2\left(\Bbb R^n\right)$,
$H$ be
essentially self adjoint, then the
following is true 
$$\align
  {}&\left[\text{exp}\left(\dfrac{-it\bar H}{\hbar}\right)
  \psi\right]\left(\vec x\right) = 
  \lim_{\eta ,\gamma\to 0}
  \int_{\Bbb R^{n}}
   \psi\left(\vec x_0\right)
   K\left(\vec x, \vec x_{0},\eta ,\gamma ,t\right)
   \,d\vec x_0, \tag2.6a
\endalign
$$
where the integral in (2.6a) is a Lebesgue integral
and the limits are taken independent of each other 
in the $L^2$ topology.  
\endproclaim

Part a of the fourth theorem above provides us with
another way to deal with arbitary $L^2$ wave functions when
equation 2.1 is considered as a distribution in $\Bbb R^{2n}$.
\proclaim{Fourth Theorem Part b} Let
$\psi ,\phi \in L^2\left(\Bbb R^n\right)$,
$H$ be
essentially self adjoint, then the
following is true
$$\align
  {}&\int_{\Bbb R^n}\phi\left(\vec x\right)
   \left[\text{exp}\left(\dfrac{-it\bar H}{\hbar}\right)
  \psi\right]\left(\vec x\right) = \\
  {}&\lim_{\eta ,\gamma\to 0}
     \int_{\Bbb R^{n}}
       \phi\left(\vec x\right)\Bigg(
        \int_{\Bbb R^{n}}
   \psi\left(\vec x_0\right)
   K\left(\vec x, \vec x_{0},\eta ,\gamma ,t\right)
   \,d\vec x_0\Bigg) d\vec x \tag2.6b
\endalign
$$
where the integrals in (2.6b) are iterated Lebesgue integrals
and the limits are taken independent of each other and 
pointwise in $t$.
\endproclaim

We will connect the second theorem to the theory
of distributions.   
\proclaim{Fifth Theorem} Let S be the space of  
rapidly decreasing 
test functions.  Suppose $H = H_0 + V$
is essentially self-adjoint,
then there exists a tempered distribution
$K_t\left(\vec x , \vec x_0\right)$ on
$S\left(\Bbb R^n \times \Bbb R^n\right)$ such that
$\forall
\phi\left(\vec x\right),
\psi\left(\vec x_0\right) \in S\left(\Bbb R^n\right)
$,
$$
\int_{\Bbb R^n} \phi\left(\vec x\right)\left[
\text{exp}\left(\dfrac{-it\bar H}{\hbar}\right)\psi\right]
\left(\vec x\right)
d\vec x\,\, = \,\,
\int K_t\left(\vec x, \vec x_0\right)
\phi\left(\vec x\right)\psi\left(\vec x_0\right)
d\vec x d\vec x_0, \tag2.7
$$
where the integral in the right hand side
of equation (2.7) is a distribution inner product.
\endproclaim
The second theorem above is linked to the theory of
tempered distributions via the fifth theorem.  
We will prove some properties of the
distribution $K_t\left(\vec x, \vec x_0\right)$
given in (2.7).  We have
in fact gone beyond the theory of distributions in the sense
that the second, and fourth theorem are not just
true for rapidly decreasing test functions, 
they are true for a much bigger class
of functions.  

Finally, the above theorems can be cast into the language
of Nonstandard Analysis.  The idea of using Nonstandard
Analysis to formulate the Feynman path integral is not new
(see [9],[13],[18],[19]
and references within).  The usual formulation is to replace the time
slice limit with a standard part.  Following that idea,
the third theorem can be easily reformulated in the language
of Nonstandard Analysis as follows.

\proclaim{Sixth Theorem} With the notations and conditions
in the third
theorem, we can write
$$
  K\left(\vec x, \vec x_{0},\eta ,\gamma ,t\right) = 
  \lim_{k\to\infty}
  K_k\left(\vec x, \vec x_{0},\eta ,\gamma ,t\right),\tag2.8 
$$
where $K_k$ is given in 2.5.  
Let $\omega\in {}^*\Bbb N - \Bbb N$, then
$$
  K\left(\vec x, \vec x_{0},\eta ,\gamma ,t\right) = 
  st\left({}^*K_\omega\left(\vec x, \vec x_{0},\eta ,\gamma ,t\right)
   \right),\tag2.9 
$$
where $st$ is the standard part.   
\endproclaim
Notice that ${}^*K_\omega$ consists of an infinite(namely
$\omega$) copies of *-improper Riemann integrals.
Equation 2.9 is in fact a special case of the work done
in reference [9].

We now  come to a new formulation of the Feynman path integral
in the language of Nonstandard Analysis.
\proclaim{Seventh Theorem} Under the conditions of the second
theorem(part a or part b), 
let $\eta$ and $\gamma$ be positive infinitesimals in the
language of Nonstandard Analysis, then
$$\align
  {}&\int_{\Bbb R^n} \phi\left(\vec x\right)\left[  \text{exp}
  \left(\dfrac{-it\bar H}{\hbar}\right)\psi\right]\left(\vec x\right)
  d\vec x  = \tag2.10a\\
  {}& st\left({}^*\bar{\int}_{\Bbb R^{2n}}\phi\left(\vec x\right)
   \psi\left(\vec x_0\right)
   K\left(\vec x, \vec x_{0},\eta ,\gamma ,t\right)
   \,d\vec x_0 d\vec x\right),
\endalign
$$
$$\align
  {}&\int_{\Bbb R^n} \phi\left(\vec x\right)\left[  \text{exp}
  \left(\dfrac{-it\bar H}{\hbar}\right)\psi\right]\left(\vec x\right)
  d\vec x  = \tag2.10b\\
  {}& st\left({}^*\int_{\Bbb R^{2n}}\phi\left(\vec x\right)
   \psi\left(\vec x_0\right)
   K\left(\vec x, \vec x_{0},\eta ,\gamma ,t\right)
   \,d\vec x_0 d\vec x\right),
\endalign
$$

where equations 2.10a and 2.10.b corresponds to theorem
two part a and part b respectively, $st$  means 
the standard part of the *-transformed
improper Lebesgue integral(2.10a, in the sense in second theorem
part a) and Lebesgue integral(2.10b, second theorem part b).  
\endproclaim
The implication of the seventh
theorem on Feynman path integrals is the following.  
Suppose the Hamiltonian has a finite number of singularities
and discontinuities, then the third theorem holds.
*-transforming the third theorem, equation 2.5 reads:
for all $\eta, \gamma\in {}^*\Bbb R^{+}$,
$$
 {}^*K\left(\vec x, \vec x_{0},\eta ,\gamma ,t\right) = 
 {}^*\lim_{k\to\infty}
    {}^*K_k\left(\vec x, \vec x_{0},\eta ,\gamma ,t\right), 
  \tag2.11
$$
where the *-limit is a limit taken in the nonstandard world,
and ${}^*K_k$ is $k$ copies of *-improper Riemann integrals.
In particular, we can let $\eta$ and $\gamma$ be positive
infinitesimals and use equation 2.11 in the seventh theorem.
The time sliced Feynman path integral now has time sliced
limits taken in the nonstandard world and standard parts
taken on $\eta$ and $\gamma$ in the sense of the seventh
theorem.  As mentioned earlier, this formulation differs from
the popular usage of Nonstandard analysis on path integrals 
in that the time slice limit is not replaced by a standard
part.  This new formulation uses the fact that the functions
$F$ and $G$ behave like delta functions when
$\eta$ and $\gamma$ are positive infinitesimals.

Lastly, we will use equation 2.9 and theorem 2 part b to compute
the harmonic oscillator propagator.  We will compute the
propagator in such a way that no prior knowledge of
the classical path is needed.  In fact, the classical
part of the propagator naturally falls out from
quantum considerations.  The usual method to compute
the harmonic oscillator with the Feynman path integral
is to use the classical path and seperate out the
classical and quantum fluctuation parts(see [4] and [5]).
From our computational point of view, 
the classical mechanics part comes purely 
from quantum considerations and that goes against the grain
of Feynman's original idea that quantum mechanics come from
classical mechanics via the action integral in the integrand
of integration over path space.  

The Hamiltonian for
the harmonic oscillator is  
$H = \dfrac{-\hbar^2}{2m}\Delta + 
\dfrac{m\lambda^{2}}{2}\vec x^2$.  
It is well known that
for $0 < t < \dfrac{\pi}{\lambda}$, the $n$ dimensional 
harmonic oscillator
propagator is given by
$$\align
{}&K\left(\vec x, \vec x_0,t\right) = \tag2.12\\
{}&\left(\dfrac{m}{2\pi{}i\hbar}\right)^{\frac{n}2}
\left(\dfrac{\lambda}
{\sin\lambda{}t}\right)^{\frac{n}{2}}
\text{exp}\left\{\dfrac{im}{\hbar}\dfrac{\lambda}{\sin\lambda{}t}
\left[\left({\vec x_0}^2 + \vec x^2\right)\cos\lambda{}t - 
2\vec x\vec x_0\right]
\right\} = \\
{}&h\left(\dfrac{n}{2}, 
t\right)g\left(\vec x, \vec x_0, t\right), \\
{}&h\left(\dfrac{n}{2}, t\right) = 
   \left(\dfrac{m}{2\pi{}i\hbar}\right)^{\frac{n}2}
\left(\dfrac{\lambda}
{\sin\lambda{}t}\right)^{\frac{n}{2}}, \\
{}&g\left(\vec x, \vec x_0, t\right) = 
\text{exp}\left\{\dfrac{im}{\hbar}\dfrac{\lambda}{\sin\lambda{}t}
\left[\left({\vec x_0}^2 + \vec x^2\right)\cos\lambda{}t -
2\vec x\vec x_0\right]
\right\}
\endalign$$
where the $g\left(\vec x, \vec x_0, t\right)$ is
the classical part and the $h\left(\dfrac{n}{2}\right)$
is the quantum fluctuation.  Given 2.12, we would expect
that 
$$
  K\left(\vec x, \vec x_{0},\eta ,\gamma ,t\right) = 
  h\left(\dfrac{n}{2}, t\right)\int_{\Bbb R^{2n}}
  g\left(\vec y, \vec y_0, t\right)F\left(\vec y_0, 
  \vec x_0, \gamma\right) 
  G\left(\vec y, \vec x, \eta\right)d\vec y d\vec y_0.\tag2.13 
$$ 
Notice that in 2.13, the disturbance of the functions $F$
and $G$ affects only the classical part.   

We conclude this section with a comment and a summary.  The reader
should compare the similarities and diferences between
the formulations above and that of the notion
of weak integral kernels   
(see [2] and references
within). In particular, one difference is that
the above kernel 
exists for all essentially self-adjoint
Hamiltonians.  
The main purpose of this paper is to 
derive a rigorous theory of real time
propagators and real time Feynman path integrals.
The propagator exists for all essentially
self-adjoint Hamiltonians and is closely 
related to distributions.  The Feynman
path integral exists for potentials that
carry at most a finite number of singularities
and discontinuities, it is formulated via 
improper Riemann integrals, and it can be
formulated with classical analysis 
and Nonstandard Analysis.  Lastly, we use
Nonstandard Analysis and 
compute the propagator for the harmonic oscillator
without prior knowledge of the classical path.   

\subhead
{\bf III. Proof of Third Theorem}\endsubhead
We start by giving a quick proof 
of the third theorem.  
The third theorem is a specific case of the work
work done in reference [9].  For the full details
of the proof, we refer the reader to [9].

We first set some notations.  
Suppose $V$ is such
that it has at most a finite
number of singularities and discontinuities.  
Let $k\in \Bbb N$ and $1\leq l\leq k+1$.  We will
denote the interior of the $l$th box by  
$$ 
  A^l = (-a_1^l, b_1^l)\times\dots\times (-a_n^l, b_n^l),\tag3.1
$$
for positive and large $a$'s and $b$'s.  
Let $K = \left\{\vec y_1\dots\vec y_p\right\}$
be the set of discontinuous and singular
points of $V$.  For each
$\vec y_q = (y_1^q,\dots ,y_n^q)\in K$, denote
the $l$th box centered at $\vec y_q$ by 
$$
  B_q^l = (y_1^q - \frac{1}{c_1^{q,l}},\,  
   y_1^q + \frac{1}{d_1^{q,l}})\times\dots\times
   (y_n^q - \frac{1}{c_n^{q,l}},\,  
   y_n^q + \frac{1}{d_n^{q,l}}), \tag3.2
$$
for positive and large $c$'s and $d$'s.
Let 
$$
  C^l = A^l - \left\{\bigcup_{q = 1}^{p}B_q^l\right\}.\tag3.3
$$
For arbitrary large $a$'s, $b$'s, 
$c$'s, and $d$'s, $C^l$ is a box which encloses
the set $K$ and at each point of $K$, a small
box centered at that point is taken out.  
Associated with $C^l$ is a set of indices 
$$\align
   \left\{j_l\right\} = \{a_1^l,\dots ,a_n^l, b_1^l,\dots ,b_n^l,
    {}&c_1^{1,l},\dots ,c_n^{1,l}, \dots ,
    c_1^{p,l},\dots ,c_n^{p,l}, \tag3.4\\
    {}&d_1^{1,l},\dots ,d_n^{1,l}, \dots ,
    d_1^{p,l},\dots ,d_n^{p,l}\}
\endalign
$$
We will denote by $\left\{j_l\right\}\to\infty$ to
mean
$$\align
    {}&a_1^l,\dots ,a_n^l, b_1^l,\dots ,b_n^l,
    c_1^{1,l},\dots ,c_n^{1,l}, \dots ,
    c_1^{p,l},\dots ,c_n^{p,l},\tag3.5\\
    {}&d_1^{1,l},\dots ,d_n^{1,l}, \dots ,
    d_1^{p,l},\dots ,d_n^{p,l}\to\infty ,
\endalign
$$
where all indices goes to infinity independent
of each other.  Notice that as 
$\left\{j_l\right\}\to\infty$,
we recover $\Bbb R^n$ $a.e.$ from $C^l$.  
We will denote by $\chi_{\left\{j_l\right\}}$ the characteristic
function on $C^l$.  Notice that 
for $f\in L^2\left(\Bbb R^n\right)$,
$$
 \lim_{\left\{j_l\right\}\to\infty}
 \chi_{\left\{j_l\right\}}f = f \quad a.e. ,\tag3.6
$$
where the limit in (3.6) is taken in the $L^2$ topology.
Let us write
$$
  D_{\left\{J_l^h\right\}} = 
  C^l\times\dots\times C^h, \quad l\leq h. \tag3.7
$$
Associated with $D_{\left\{J_l^h\right\}}$
is a set of indices
$$
  \left\{J_l^h\right\} = 
  \bigcup_{\alpha = l}^{h}\left\{j_\alpha\right\}, \tag3.8
$$
and as before, we will use the notation
$\left\{J_l^h\right\}\to\infty$ to mean
$$
  \left\{j_l\right\}\to\infty ,\dots 
  ,\left\{j_h\right\}\to\infty , \tag3.9
$$
where the indices are taken to infinity
independent of each other.
Finally, we will denote by $\int_{rO}$ to be
Riemann or improper Riemann integration over
the region $O$ and $\int_{O}$ to be
Lebesgue integration over the region $O$.

\proclaim{\bf Theorem 3.1} The third theorem in
  section II is true.
\endproclaim

\demo{Proof} Trotter's product formula(see [7],[8], and [10])
and Schwarz's inequality
implies that
$$\align
  {}&\int_{\Bbb R^n} G_{\vec x}\left(\vec y,\eta\right)\left[
  \text{exp}
  \left(\dfrac{-it\bar H}{\hbar}\right)
  F_{\vec x_0}\left(\vec z, \gamma\right)
  \right]\left(\vec y\right)
  d\vec y  = \tag3.10\\
  {}&\lim_{k\to\infty}
     \int_{\Bbb R^n} G_{\vec x}\left(\vec y,\eta\right)
  \left[\left\{\text{exp}\left(\dfrac{-itV}{k\hbar}\right)
  \text{exp}\left(\dfrac{-itH_0}{k\hbar}\right)
  \right\}^k
  F_{\vec x_0}\left(\vec z, \gamma\right)
  \right]\left(\vec y\right)\,d\vec y\,  ,
\endalign
$$
where the limit in (3.10) is taken pointwise as a function
of $t$.  To the right of each of the operator
$\text{exp}\left(\dfrac{-itH_0}{k\hbar}\right)$
in (3.10), we put in the identity operator
$\lim_{\left\{j_l\right\}\to\infty}
  \chi_{\left\{j_l\right\}}$ for $1 \leq l\leq k$
in increasing order from right to left and the 
limit is taken in the $L^2$ topology.  Since
$\text{exp}\left(\dfrac{-itV}{k\hbar}\right)
$, $\text{exp}\left(\dfrac{-itH_0}{k\hbar}\right)$,
and multiplication by a characteristic function
are all continuous operators, we can take all the limits
outside of the operators and get
$$\align
  {}&\left\{\text{exp}\left(\dfrac{-itV}{k\hbar}\right)
  \text{exp}\left(\dfrac{-itH_0}{k\hbar}\right)
  \right\}^kF_{\vec x_0} =  \tag3.11\\
  {}&\text{exp}\left(\dfrac{-itV}{k\hbar}\right)
  \text{exp}\left(\dfrac{-itH_0}{k\hbar}\right)
  \lim_{\left\{j_{k}\right\}\to\infty}
  \chi_{\left\{j_{k}\right\}}\dots \\
  {}&\text{exp}\left(\dfrac{-itV}{k\hbar}\right)
  \text{exp}\left(\dfrac{-itH_0}{k\hbar}\right)
  \lim_{\left\{j_{1}\right\}\to\infty}
  \chi_{\left\{j_{1}\right\}}F_{\vec x_0} = \\
  {}&\lim_{\left\{J_1^{k}\right\}\to\infty}
  \text{exp}\left(\dfrac{-itV}{k\hbar}\right)
  \text{exp}\left(\dfrac{-itH_0}{k\hbar}\right)
  \chi_{\left\{j_{k}\right\}}\dots \\
  {}&\text{exp}\left(\dfrac{-itV}{k\hbar}\right)
  \text{exp}\left(\dfrac{-itH_0}{k\hbar}\right)
  \chi_{\left\{j_{1}\right\}}F_{\vec x_0} = \\
  {}&\lim_{\left\{J_1^{k}\right\}\to\infty}
    w_{n,k}
    \int_{D_{\left\{J_1^{k}\right\}}}
    \text{exp}\left[\dfrac{i\epsilon}{\hbar}
     S_k\left(\vec x_{k+1}, \dots ,\vec x_1\right)
     \right]F_{\vec x_0}\left(\vec x_1, \gamma\right)
     \,d\vec x_1\dots d\vec x_{k}.
\endalign 
$$
In the last equality in (3.11), we
used the integral representation of
the free evolution operator; we emphasize that
all limits in (3.11) are taken in the $L^2$ 
topology and are taken independent of
each other.  
Equations (3.6), (3.10), (3.11) and 
Schwarz's inequality imply that
$$\align
  {}&\int_{\Bbb R^n} G_{\vec x}\left(\vec y,\eta\right)\left[
  \text{exp}
  \left(\dfrac{-it\bar H}{\hbar}\right)
  F_{\vec x_0}\left(\vec z, \gamma\right)
  \right]\left(\vec y\right)
  d\vec y  = \tag3.12\\
  {}&\lim_{k\to\infty}w_{n,k}\int_{\Bbb R^n}\bigg[
     \lim_{\left\{j_{k+1}\right\}\to\infty}
     \chi_{\left\{j_{k+1}\right\}}
     G_{\vec x}\left(\vec x_{k+1},\eta\right)
     \times\\
  {}&\lim_{\left\{J_1^{k}\right\}\to\infty}
     w_{n,k}
    \int_{D_{\left\{J_1^{k}\right\}}}
    \text{exp}\left[\dfrac{i\epsilon}{\hbar}
     S_k\left(\vec x_{k+1}, \dots ,\vec x_1\right)
     \right]F_{\vec x_0}\left(\vec x_1, \gamma\right)
     \,d\vec x_1\dots d\vec x_{k}
     \bigg]d\vec x_{k+1} = \\
  {}&\lim_{k\to\infty}w_{n,k}
     \lim_{\left\{J_1^{k+1}\right\}\to\infty}
     \int_{D_{\left\{J_1^{k+1}\right\}}}
     G_{\vec x}\left(\vec x_{k+1},\eta\right)
     \times\\
  {}&\text{exp}\left[\dfrac{i\epsilon}{\hbar}
     S_k\left(\vec x_{k+1}, \dots ,\vec x_1\right)\right]
     F_{\vec x_0}\left(\vec x_{1}, \gamma\right)
     \,d\vec x_1\dots d\vec x_{k+1}.
\endalign
$$
In (3.12), all limits inside the integrals are taken 
independent of each other in 
the $L^2$ topology and all limits taken outside of 
the integral are taken pointwise in $t$.  
By construction, the integrand
$$     G_{\vec x}\left(\vec x_{k+1},\eta\right)
     \text{exp}\left[\dfrac{i\epsilon}{\hbar}
     S_k\left(\vec x_{k+1}, \dots ,\vec x_1\right)\right]
     F_{\vec x_0}\left(\vec x_{1}, \gamma\right)\tag3.13
$$
is bounded and continuous on $D_{\left\{J_1^{k+1}\right\}}$.
Hence, the Lebesgue integral over $D_{\left\{J_1^{k+1}\right\}}$
in the last equality of (3.12)
can be replaced by a Riemann integral over
$D_{\left\{J_1^{k+1}\right\}}$.  Since the
$\left\{J_1^{k+1}\right\}$ limits in (3.12) are all taken
independent of each other, we can interpret those
limits and the integral as an improper Riemann integral.  \qed
\enddemo
\remark{\bf Remark 3.2}
It is not necessary to use improper Riemann integrals or
put the discontinuities and singularities restriction on the 
potential.  We forget
about the holes centered at each elements of 
$K$ as given in (3.2) and take $C^l = A^l$ as defined
in (3.1).
Proceeding in the same manner as the above proof, we get to (3.12).
At this point, we do not replace the Lebesgue integral with
Riemann integral since the integrand is not necessarily bounded
and continuous over the region of integration.  We are then left
with an improper Lebesgue integral in which the convergence
of the integral is taken pointwise in $t$.  
\endremark
\subhead
{\bf IV. Proof of First Theorem}
\endsubhead
In this section, we prove the first theorem in 
section II.
Let us denote
$$\align
 &{}T^{k} = \left\{\text{exp}
  \left(\dfrac{-itV}{k\hbar}\right)
  \text{exp}\left(\dfrac{-itH_0}{k\hbar}\right)
  \right\}^{k}, \tag4.1\\
 &{}\bar T^{k} = \left\{
  \text{exp}\left(\dfrac{-itH_0}{k\hbar}\right)
  \text{exp}
  \left(\dfrac{-itV}{k\hbar}\right)
  \right\}^{k}.
\endalign
$$

\proclaim{Theorem 4.1}With our previously defined
notations, we have 
$$\align
  {}&\big|K\left(\vec x, \vec x_0,
      \eta ,\gamma ,t\right)|
      \leq C_{t,\eta ,\gamma} \tag4.2
\endalign
$$
where $C_{t,\eta ,\gamma}$ 
is a constant depending only on $t, \eta$,
and $\gamma$.
\endproclaim
\demo{Proof}Since the evolution operator has norm 1,
using
Schwarz's inequality on the kernel in 2.1 gives
$$\align
  {}&|K\left(\vec x, \vec x_0,
      \eta ,\gamma ,t\right)|
      \leq  \tag4.3\\
  {}&\big|\big|G_{\vec x}\left(\vec y, \eta\right)
     \big|\big|_2 
  \big|\big|
  F_{\vec x_0}
  \left(\vec y, \gamma\right)\big|\big|_2  
      \equiv C_{t,\eta ,\gamma}.  \qed 
\endalign
$$
\enddemo

We will now show that  
$K\left(\vec x, \vec x_0,
      \eta ,\gamma ,t,
      \right)$ 
is continuous as functions of $\left(\vec x,\vec x_0\right)
\in\Bbb R^n$. 

\proclaim{Lemma 4.2} Let $f,g\in L^2\left(\Bbb R^n\right)$, 
then the following is true 
$$\align
  {}&\int_{\Bbb R^n}g\left(\vec x\right)
  \left[\text{exp}\left(\dfrac{-itH_0}{k\hbar}\right)
  f\right]\left(\vec x\right)d\vec x = 
  \int_{\Bbb R^n}
  \left[\text{exp}\left(\dfrac{-itH_0}{k\hbar}\right)
  g\right]\left(\vec x\right) 
  f\left(\vec x\right)
  d\vec x .\tag4.4 
\endalign
$$
\endproclaim
\demo{Proof} Let $\chi_\alpha$ be the characteristic function
of the cube centered at the origin with sides of length
$\alpha$, then 
$$\align
  {}&\left[\text{exp}\left(\dfrac{-i\epsilon{} H_0}{\hbar}\right)
  f\right]\left(\vec x\right) = \tag4.5 \\
  {}&\lim_{\alpha\to\infty}w_{n,1}
  \int_{\Bbb R^n} \chi_{\alpha}\left(\vec y\right)
  \text{ exp }\left[\dfrac{im\epsilon}{2\hbar}
  \left(\dfrac{\vec x - \vec y}{\epsilon}\right)^2
  \right]
  f\left(\vec y \right)\, d\vec y , 
\endalign
$$
where the limit in (4.5) is taken in the $L^2$ norm.
Using Schwarz's inequality on $\alpha$ and
Lebesgue's dominating convergence theorem on
$\beta$, we have
$$\align
  {}&\int_{\Bbb R^n}g\left(\vec x\right)
  \left[\text{exp}\left(\dfrac{-itH_0}{k\hbar}\right)
  f\right]\left(\vec x\right)d\vec x = \tag4.6\\
  {}&\lim_{\beta ,\alpha\to\infty}w_{n,1}
  \int_{\Bbb R^n}\chi_{\beta}\left(\vec x\right)
  g\left(\vec x\right)\Bigg\{
  \int_{\Bbb R^n} \chi_{\alpha}\left(\vec y\right)
  \text{ exp }\left[\dfrac{im\epsilon}{2\hbar}
  \left(\dfrac{\vec x - \vec y}{\epsilon}\right)^2
  \right]
  f\left(\vec y \right)\, d\vec y\Bigg\} d\vec x =\\
  {}&\lim_{\beta ,\alpha\to\infty}w_{n,1}
  \int_{\Bbb R^n}\chi_{\alpha}\left(\vec y\right)
  f\left(\vec y \right)
  \Bigg\{
  \int_{\Bbb R^n} 
  \chi_{\beta}\left(\vec x\right)
  \text{ exp }\left[\dfrac{im\epsilon}{2\hbar}
  \left(\dfrac{\vec x - \vec y}{\epsilon}\right)^2
  \right]g\left(\vec x\right)
  \, d\vec x\Bigg\} d\vec y ,
\endalign
$$
where the limits are taken pointwise in $t$.
Using Schwarz's inequality on $\beta$ and
Lebesgue's dominating convergence theorem
on $\alpha$ in the last expression in (4.6)
gives (4.4). \qed
\enddemo

\proclaim{Lemma 4.3} Let $f,g\in L^2\left(\Bbb R^n\right)$,
then the following is true
$$\align
  {}&\int_{\Bbb R^n}g\left(\vec x\right)
  \left[\text{exp}\left(\dfrac{-it\bar H}{\hbar}\right)
  f\right]\left(\vec x\right)d\vec x =
  \int_{\Bbb R^n}
  \left[\text{exp}\left(\dfrac{-it\bar H}{\hbar}\right)
  g\right]\left(\vec x\right)
  f\left(\vec x\right)
  d\vec x .\tag4.7
\endalign
$$
\endproclaim

\demo{Proof} Intuitively, if we think of the evolution 
as an exponential
of the Hamiltonian $\bar H$, we can expand the exponential
in powers of $\bar H$ and put all powers of $\bar H$ from
the function $f$ onto the function $g$ since $\bar H$ is
self-adjoint.  This is also true for lemma 4.2.

Schwarz's inequality and Trotter's formula
implies
$$\align
  {}&\int_{\Bbb R^n}g\left(\vec x\right)
  \left[\text{exp}\left(\dfrac{-it\bar H}{\hbar}\right)
  f\right]\left(\vec x\right)d\vec x =
  \lim_{k\to\infty}
   \int_{\Bbb R^n}g\left(\vec x\right)
   \left[T^kf\right]\left(\vec x\right)d\vec x,\tag4.8
\endalign
$$
where the limit is taken pointwise in $t$ and
$T^k$ is given in (4.1).  Using Lemma 4.2,
and $\bar T^k$ as defined in (4.1), we obtain
$$\align
  {}&\lim_{k\to\infty}
   \int_{\Bbb R^n}g\left(\vec x\right)
   \left[T^kf\right]\left(\vec x\right)d\vec x = 
  \lim_{k\to\infty}
   \int_{\Bbb R^n}
   \left[\bar T^kg\right]\left(\vec x\right)  
   f\left(\vec x\right)d\vec x = \\
  {}&\int_{\Bbb R^n}
  \left[\text{exp}\left(\dfrac{-it\bar H}{\hbar}\right)
  g\right]\left(\vec x\right)
  f\left(\vec x\right)
  d\vec x .\qed \tag4.9
\endalign
$$
\enddemo

\proclaim{Theorem 4.4} With our previously defined notations,
the expression 
$$\align
  {}&\bigg|K\left(\vec x, \vec x_0,
      \eta ,\gamma ,t 
      \right) - 
     K\left(\vec y, \vec y_0,
      \eta ,\gamma ,t
      \right)\bigg| \tag4.10
\endalign
$$
goes to zero as $\left(\vec x, \vec x_0\right)$ goes to
$\left(\vec y, \vec y_0\right)$.
\endproclaim
\demo{Proof}We first show that 
$K\left(\vec x, \vec x_0,
      \eta ,\gamma ,t
      \right)$
is seperately continuous in $\vec x$
and $\vec x_{0}$, then jointly continuous.
Schwarz's inequality implies that
$$\align
  {}&\big|K\left(\vec x, \vec x_0,
      \eta ,\gamma ,t 
      \right) -
  K\left(\vec y, \vec x_0,
      \eta ,\gamma ,t 
      \right)\big|^2 \leq \tag4.11 \\
  {}&\big|\big|G_{\vec x}\left(\vec z, \eta\right) -
    G_{\vec y}\left(\vec z, \eta\right)\big|\big|_2^2 
    \times \big|\big|
      F_{\vec x_0}
       \left(\vec z, \gamma\right)\big|\big|_2^2 = \\
  {}&\big|\big|F_{\vec x_0}\left(\vec z, \gamma\right)
    \big|\big|_2^2 \times 
    \left(\dfrac{m}{2\pi\hbar\eta}\right)^
    {n}\int_{\Bbb R^n}\Bigg\{
       \text{ exp }\left[\dfrac{-m\eta}{2\hbar}
  \left(\dfrac{\vec x - \vec z}{\eta}\right)^2
  \right] - \\
  {}&\text{ exp }\left[\dfrac{-m\eta}{2\hbar}
  \left(\dfrac{\vec y - \vec z}{\eta}\right)^2
  \right]\Bigg\}^2 d\vec z = 
  C\left(\dfrac{m}{2\pi\hbar\eta}\right)^
    {n}\Bigg\{
    \int_{\Bbb R^n}2\left[
    \text{ exp }\left(\dfrac{-m\eta}{2\hbar}
    \dfrac{\vec z^2}{\eta^2}\right)\right]^2d\vec z - \\
  {}&\int_{\Bbb R^n}2\text{ exp }\left(\dfrac{-m\eta}{2\hbar}
    \dfrac{\vec z^2}{\eta^2}\right)
    \text{ exp }\left(\dfrac{-m\eta}{2\hbar}
    \dfrac{\left(\vec z +
    \vec y - \vec x\right)^2}
    {\eta^2}\right)
    d\vec z\Bigg\} = C_{\gamma ,t}
    g\left(\vec y, \vec x\right),
\endalign
$$
where $C_{\gamma ,t}$ is a constant independent of
$\vec x_0$ and $g\left(\vec y, \vec x\right)$ 
is independent of $\vec x_0$.
Using Lebesgue's dominating convergence theorem
on $\vec x\to\vec y$
in (4.11), we get $C_{\gamma ,t}
g\left(\vec y, \vec x\right)\to 0$.
Using lemma 4.3 in equation 2.1, 
we can put the evolution operator
on $G_{\vec x}\left(\vec y ,\eta\right)$.  With
the same reasoning as 4.11, we get
$$\align
  {}&\big|K\left(\vec x, \vec x_0,
      \eta ,\gamma ,t
      \right) -
  K\left(\vec x, \vec y_0,
      \eta ,\gamma ,t
      \right)\big|^2 \leq \tag4.12 \\
  {}&D_{\eta ,t}f\left(\vec y_0, \vec x_0\right)\to 0
\endalign
$$
as $\vec x_0\to\vec y_0$.

Finally,
$$\align
  {}&\bigg|K\left(\vec x, \vec x_0,
      \eta ,\gamma ,t 
      \right) -
     K\left(\vec y, \vec y_0,
      \eta ,\gamma ,t 
      \right)\bigg| \leq\tag4.13 \\
   {}&\bigg|K\left(\vec x, \vec x_0,
      \eta ,\gamma ,t 
       \right) -
      K\left(\vec y, \vec x_0,
      \eta ,\gamma ,t
      \right)\bigg| + \\
   {}&\bigg|K\left(\vec y, \vec x_0,
      \eta ,\gamma ,t 
      \right) -
      K\left(\vec y, \vec y_0,
      \eta ,\gamma ,t 
      \right)\bigg| \leq \\
   {}&\sqrt{C_{\gamma ,t}g\left(\vec y, \vec x\right)} +
      \sqrt{D_{\eta ,t}f\left(\vec y_{0}, \vec x_{0}\right)} \to 0
\endalign
$$
as $\left(\vec x, \vec x_{0}\right)
\to\left(\vec y, \vec y_{0}\right)$.
\qed
\enddemo

\subhead
{\bf V. Proof of Second, and Fourth theorem}\endsubhead
We will now prove the second and fourth theorem
in section II.  
\proclaim{\bf Proposition 5.1} Let $f,g\in L^2$, then
$$\align
   {}&\int_{\Bbb R^n}f\left(\vec z\right)
   \left[\text{exp}\left(\dfrac{-\eta H_0}{\hbar}\right)
   g\right]\left(\vec z\right)d\vec z = \tag5.1\\
   {}&\int_{\Bbb R^n}g\left(\vec z\right)
   \left[\text{exp}\left(\dfrac{-\eta H_0}{\hbar}\right)
   f\right]\left(\vec z\right)d\vec z = \\
   {}&\int_{\Bbb R^{2n}}g\left(\vec x\right)
     G\left(\vec x, \vec y, \eta\right)
     f\left(\vec y\right)d\vec x d\vec y,
\endalign
$$
where $G\left(\vec x, \vec y, \eta\right)$ is the
Gaussian kernel given in equation 2.1.
\endproclaim
\demo{Proof} Since $|f|$ and $|g|$ are in $L^2$, we have
$$\align
  {}&\int_{R^n}
      \Bigg(\int_{R^n}
      |g\left(\vec x\right)|
      |G\left(\vec x, \vec y, \eta\right)|
      |f\left(\vec y\right)|d\vec x\Bigg)d\vec y = \tag5.2\\
  {}&\int_{R^n}
      |g\left(\vec x\right)|
       \Bigg(\int_{R^n}
      G\left(\vec x, \vec y, \eta\right)
      |f\left(\vec y\right)|d\vec x\Bigg)d\vec y = \\
  {}&\int_{\Bbb R^n}|g\left(\vec x\right)|
   \left[\text{exp}\left(\dfrac{-\eta H_0}{\hbar}\right)
    |f|\right]\left(\vec x\right)d\vec x < \infty.
\endalign
$$
Equation 5.1 then follows from Fubini's theorem.  \qed
\enddemo

\proclaim{\bf Theorem 5.2} Part b of the second theorem in 
section II is true.  
\endproclaim
\demo{Proof} First recall that in part b of the second
theorem, the wave functions 
$\phi, \psi \in L^2\cap L^1$.  
Using lemma 4.3,
the kernel in (2.1) can be written as
$$\align
   {}&K\left(\vec x, \vec x_{0},\eta ,\gamma ,t\right) =
  \left[\text{exp}
  \left(\dfrac{-\gamma H_0}{\hbar}\right)
  \text{exp}
  \left(\dfrac{-it\bar H}{\hbar}\right)
   G_{\vec x}\left(\vec z, \eta\right)
   \right]\left(\vec x_0\right).
   \tag5.3\\
\endalign
$$
We have that
$$\align
  {}&\left[\text{exp}\left(\dfrac{-it\bar H}{\hbar}\right)
   \psi\right]\left(\vec x\right) = \tag5.4\\
  {}&\lim_{\eta ,\gamma\to 0}
  \left[\text{exp}\left(\dfrac{-\eta H_0}{\hbar}\right)
  \text{exp}\left(\dfrac{-it\bar H}{\hbar}\right)
   \text{exp}\left(\dfrac{-\gamma H_0}{\hbar}\right)
   \psi\right]\left(\vec x\right),
\endalign
$$
where the limits in 5.4 are taken in $L^2$ topology. 
Using Schwarz's inequality, the following holds
$$\align
  {}&\int_{\Bbb R^n}\phi\left(\vec x\right)
   \left[\text{exp}\left(\dfrac{-it\bar H}{\hbar}\right)
   \psi\right]\left(\vec x\right)d\vec x = \tag5.5\\
  {}&\lim_{\eta ,\gamma\to 0}
    \int_{\Bbb R^n}\phi\left(\vec x\right)
  \left[\text{exp}\left(\dfrac{-\eta H_0}{\hbar}\right)
  \text{exp}\left(\dfrac{-it\bar H}{\hbar}\right)
   \text{exp}\left(\dfrac{-\gamma H_0}{\hbar}\right)
   \psi\right]\left(\vec x\right)d\vec x,
\endalign
$$
where the limits in 5.5 are taken pointwise in $t$.
Lemma 4.3 and proposition 5.1 implies that
$$\align
  {}&\left[\text{exp}\left(\dfrac{-\eta H_0}{\hbar}\right)
  \text{exp}\left(\dfrac{-it\bar H}{\hbar}\right)
   \text{exp}\left(\dfrac{-\gamma H_0}{\hbar}\right)
   \psi\right]\left(\vec x\right) = \tag5.6\\
  {}&\int_{\Bbb R^n} G\left(\vec x,\vec y, \eta\right)
      \left[\text{exp}\left(\dfrac{-it\bar H}{\hbar}\right)
   \text{exp}\left(\dfrac{-\gamma H_0}{\hbar}\right)
   \psi\right]\left(\vec y\right)d\vec y = \\
 {}&\int_{\Bbb R^n}
    \left[\text{exp}\left(\dfrac{-it\bar H}{\hbar}\right) 
     G_{\vec x}\left(\vec z, \eta\right)\right]\left(
     \vec y\right)\left[
     \text{exp}\left(\dfrac{-\gamma H_0}{\hbar}\right)
   \psi\right]\left(\vec y\right)d\vec y = \\
 {}&\int_{\Bbb R^n}
     \left[\text{exp}
  \left(\dfrac{-\gamma H_0}{\hbar}\right)
  \text{exp}
  \left(\dfrac{-it\bar H}{\hbar}\right)
   G_{\vec x}\left(\vec z, \eta\right)
   \right]\left(\vec x_0\right)
   \psi\left(\vec x_0\right)d\vec x_0 = \\
  {}&\int_{\Bbb R^n}
   K\left(\vec x, \vec x_{0},\eta ,\gamma ,t\right)
   \psi\left(\vec x_0\right)d\vec x_0.
\endalign
$$
Thus,
$$\align
  {}&\int_{\Bbb R^n}\phi\left(\vec x\right)
   \left[\text{exp}\left(\dfrac{-it\bar H}{\hbar}\right)
   \psi\right]\left(\vec x\right)d\vec x = \tag5.7\\
  {}&\lim_{\eta ,\gamma\to 0}
    \int_{\Bbb R^n}\phi\left(\vec x\right)
    \Bigg(\int_{\Bbb R^n}
   K\left(\vec x, \vec x_{0},\eta ,\gamma ,t\right)
   \psi\left(\vec x_0\right)d\vec x_0\Bigg)d\vec x = \\
  {}&\lim_{\eta ,\gamma\to 0}
    \int_{\Bbb R^{2n}}\phi\left(\vec x\right)
    K\left(\vec x, \vec x_{0},\eta ,\gamma ,t\right)
   \psi\left(\vec x_0\right)d\vec x_0 d\vec x,
\endalign
$$  
where the last equality in 5.7 is obtained from the first
theorem in section II(theorem 4.1 and 4.2) 
and the fact that the wave functions
$\phi$ and $\psi$ are in $L^1$.  \qed
\enddemo

\proclaim{\bf Theorem 5.3}Part a of the second theorem
in section II is true.
\endproclaim
\demo{Proof}Let $C^1 = A^1, C^2 = A^2$ be as described in 
equation 3.1 and remark 3.2.  Let $\chi^1_{\left\{j_1\right\}}, 
\chi^2_{\left\{j_2\right\}}$ be
the characteristic function on the region $C^1, C^2$ 
respectively($\left\{j_1\right\}$, and 
$\left\{j_2\right\}$
are as described in equation 3.4 for $C^1$ and $C^2$).  
Let $\phi ,\psi\in L^2$, then
$$\align
  {}&\left[\text{exp}\left(\dfrac{-\eta H_0}{\hbar}\right)
  \text{exp}\left(\dfrac{-it\bar H}{\hbar}\right)
   \text{exp}\left(\dfrac{-\gamma H_0}{\hbar}\right)
   \psi\right] = \tag5.8\\
  {}&\lim_{\left\{j_1\right\}\to\infty}
   \left[\text{exp}\left(\dfrac{-\eta H_0}{\hbar}\right)
  \text{exp}\left(\dfrac{-it\bar H}{\hbar}\right)
   \text{exp}\left(\dfrac{-\gamma H_0}{\hbar}\right)
   \chi^1_{\left\{j_1\right\}}\psi\right], \\
\endalign
$$
$$\align
  {}&\phi = \lim_{\left\{j_2\right\}\to\infty}
     \chi^2_{\left\{j_1\right\}}\phi,
\endalign
$$
where all limits in 5.8 are taken in the $L^2$ 
topology.  Using 5.4, 5.5 and 5.8, we have
$$\align
  {}&\int_{\Bbb R^n}\phi\left(\vec x\right)
   \left[\text{exp}\left(\dfrac{-it\bar H}{\hbar}\right)
   \psi\right]\left(\vec x\right)d\vec x = \tag5.9\\
  {}&\lim_{\eta ,\gamma\to 0}
    \int_{\Bbb R^n}\phi\left(\vec x\right)
  \left[\text{exp}\left(\dfrac{-\eta H_0}{\hbar}\right)
  \text{exp}\left(\dfrac{-it\bar H}{\hbar}\right)
   \text{exp}\left(\dfrac{-\gamma H_0}{\hbar}\right)
   \psi\right]\left(\vec x\right)d\vec x = \\
  {}&\lim_{\eta ,\gamma\to 0}
    \int_{\Bbb R^n}\lim_{\left\{j_2\right\}\to\infty}
     \chi^2_{\left\{j_1\right\}}\phi\left(\vec x\right)
     \times\\
   {}&\lim_{\left\{j_2\right\}\to\infty}
    \left[\text{exp}\left(\dfrac{-\eta H_0}{\hbar}\right)
  \text{exp}\left(\dfrac{-it\bar H}{\hbar}\right)
   \text{exp}\left(\dfrac{-\gamma H_0}{\hbar}\right)
    \chi^2_{\left\{j_2\right\}}
   \psi\right]\left(\vec x\right)d\vec x.
\endalign
$$
Finally, using Schwarz's inequality and theorem 5.2 on
the last expression in 5.9 gives
$$\align
  {}&\int_{\Bbb R^n}\phi\left(\vec x\right)
   \left[\text{exp}\left(\dfrac{-it\bar H}{\hbar}\right)
   \psi\right]\left(\vec x\right)d\vec x = \tag5.10\\
   {}&\lim_{\eta ,\gamma\to 0}
    \lim_{\left\{j_1\right\}, \left\{j_2\right\}\to\infty}
    \int_{\Bbb R^n}\chi^1_{\left\{j_1\right\}}
    \phi\left(\vec x\right)\times\\
   {}&\left[\text{exp}\left(\dfrac{-\eta H_0}{\hbar}\right)
  \text{exp}\left(\dfrac{-it\bar H}{\hbar}\right)
   \text{exp}\left(\dfrac{-\gamma H_0}{\hbar}\right)
   \chi^2_{\left\{j_2\right\}}
   \psi\right]\left(\vec x\right)d\vec x = \\
     {}&\lim_{\eta ,\gamma\to 0}
    \lim_{\left\{j_1\right\}, \left\{j_2\right\}\to\infty}
    \int_{\Bbb C^1\times C^2}\phi\left(\vec x\right)
    K\left(\vec x, \vec x_{0},\eta ,\gamma ,t\right)
   \psi\left(\vec x_0\right)d\vec x_0 d\vec x \equiv \\ 
  {}&\lim_{\eta ,\gamma\to 0}
    \bar{\int}_{\Bbb R^{2n}}\phi\left(\vec x\right)
    K\left(\vec x, \vec x_{0},\eta ,\gamma ,t\right)
   \psi\left(\vec x_0\right)d\vec x_0 d\vec x, 
\endalign
$$
where all limits outside of integrals are taken
pointwise in $t$ and $\bar{\int}$ by definition is
an improper Lebesgue integral with convergence at
infinity taken pointwise in $t$. \qed
\enddemo

\remark{\bf Remark 5.1} If we restrict $\phi ,
\psi\in L^2$ to have a finite number of singularities
and discontinuities as in the proof of the third theorem
(theorem 3.1), we can obtain an improper Riemann integral
as opposed to an improper Lebesgue integral for theorem
5.3.
\endremark

\proclaim{\bf Theorem 5.4}Part a of the fourth theorem
in section II is true.
\endproclaim
\demo{Proof} Equation 5.6 is true for all 
$\psi\in L^2$.  Taken $\lim_{\eta ,\gamma\to\infty}$
on both sides of 5.6 in the $L^2$ topology gives part a
of theorem four in section I equation 2.6a. \qed
\enddemo

\proclaim{\bf Theorem 5.6} Part b of the fourth theorem
in section II is true. 
\endproclaim
\demo{Proof} Let $\phi ,\psi\in L^2$, then Schwarz's inequality
and part a of the fourth theorem(theorem 5.4) implies
$$ \align
{}&\int_{\Bbb R^n}\phi\left(\vec x\right)
   \left[\text{exp}\left(\dfrac{-it\bar H}{\hbar}\right)
  \psi\right]\left(\vec x\right) = \tag5.11\\
  {}&\int_{\Bbb R^{n}}
       \phi\left(\vec x\right)\Bigg(
   \lim_{\eta ,\gamma\to 0}
        \int_{\Bbb R^{n}}
   \psi\left(\vec x_0\right)
   K\left(\vec x, \vec x_{0},\eta ,\gamma ,t\right)
   \,d\vec x_0\Bigg) d\vec x = \\
  {}&\lim_{\eta ,\gamma\to 0}
     \int_{\Bbb R^{n}}
       \phi\left(\vec x\right)\Bigg(
        \int_{\Bbb R^{n}}
   \psi\left(\vec x_0\right)
   K\left(\vec x, \vec x_{0},\eta ,\gamma ,t\right)
   \,d\vec x_0\Bigg) d\vec x,
\endalign
$$
where limits inside the integral are taken in $L^2$
and limits taken outside the integrals are taken
pointwise in $t$.\qed
\enddemo

\subhead
{\bf VI. Proof Fifth Theorem}\endsubhead
In this section, we prove the fifth theorem in section II and
derive some properties of the tempered distribution in the fifth 
theorem.  
Since we will
be working with tempered distributions, we will
let $\phi$ and $\psi$ be in the class of
rapidly decreasing test functions
which we will denote by $S\left(\Bbb R^n\right)$.  
If $\phi$ and $\psi$ are elements of $L^2\left(\Bbb R^n\right)$,
we can choose a sequence of test functions $\left\{\phi_l\right\}$
and $\left\{\psi_j\right\}$ such that $\phi_l \to \phi$ and
$\psi_j\to\psi$ in $L^2$.  Applying Schwarz's 
inequality 
and using the fact that the evolution operator has operator norm equal to 1,
we can write
$$
  \int_{\Bbb R^n} \phi\left(\vec x\right)\left[
  \text{exp}
  \left(\dfrac{-it\bar H}{\hbar}\right)\psi\right]\left(\vec x\right)
  d\vec x = \lim_{j,l\to\infty}
  \int_{\Bbb R^n} \phi_l\left(\vec x\right)\left[
  \text{exp}
  \left(\dfrac{-it\bar H}{\hbar}\right)\psi_j\right]\left(\vec x\right)
  d\vec x. \tag6.1
$$
Thus, by taking limits in the sense of (6.1),
we can always recover all of the $L^2$ wave functions in the
theory.
\proclaim{\bf Theorem 6.1} The fifth theorem in section I is true.
\endproclaim
\demo{Proof} 
Suppose $\left\{\phi{}_k\left(\vec x\right)\right\}\subset
S\left(\Bbb R^n\right) \text{with }
\phi{}_k\left(\vec x\right) \to
\phi\left(\vec x\right)
\text{in } S\left(\Bbb R^n\right)$, then
$$\align
{}&\bigg|\int_{\Bbb R^n}
\left[\phi\left(\vec x\right) - \phi_k\left(\vec x\right)\right]
\left[
\text{exp}\left(\dfrac{-it\bar H}{\hbar}\right)\psi\right]\left(\vec x\right)
d\vec x\,
\bigg| \leqq \tag6.2\\
{}&||\,\phi - \phi_k\,||_2 \times
||\,\text{exp}\left(\dfrac{-it\bar H}{\hbar}\right)\psi\,||_2 \to 0 .
\endalign
$$
Suppose $\left\{\psi{}_k\left(\vec x\right)\right\}\subset
S\left(\Bbb R^n\right) \text{with }
\psi{}_k\left(\vec x\right) \to
\psi\left(\vec x\right)
\text{in } S\left(\Bbb R^n\right)$, then
$$\align
{}&\bigg|\int_{\Bbb R^n}
\phi\left(\vec x\right)
\left[
\text{exp}\left(\dfrac{-it\bar H}{\hbar}\right)
\left(\psi - \psi{}_k\right)\right]\left(\vec x\right)
d\vec x\,
\bigg| \leqq \tag6.3\\
{}&||\,\phi\,||_2 \times 
||\,\text{exp}\left(\dfrac{-it\bar H}{\hbar}\right)
  \left(\psi - \psi{}_k\right)\,||_2 =
||\,\phi\,||_2 \times
||\,\left(\psi - \psi{}_k\right)\,||_2 \to 0 .
\endalign
$$
Hence the theorem follows from Schwartz's Kernel Theorem. \qed
\enddemo
\remark{\bf Remark 6.2} Notice that the second and the fifth theorem
imply that 
$$\align
  {}&\lim_{\eta ,\gamma\to\infty}
   \int_{\Bbb R^{2n}}\phi\left(\vec x\right)
   \psi\left(\vec x_0\right)
   K\left(\vec x, \vec x_{0},\eta ,\gamma ,t\right)
   \,d\vec x_0 d\vec x =
   \int K_t\left(\vec x, \vec x_0\right)
   \phi\left(\vec x\right)\psi\left(\vec x_0\right)
   d\vec x d\vec x_0, \tag6.4
\endalign
$$
\endremark 
At $t = 0$, the evolution operator becomes the identity 
operator.  In distributions language, We have the following   
\proclaim{\bf Theorem 6.3} At $t = 0$,
$K_t\left(\vec x, \vec x_0\right)$ satisfies
$K_0\left(\vec x, \vec x_0\right) = \delta\left(\vec x - \vec x_0\right)$
\endproclaim
\demo{Proof } Since $\left[
\text{exp}\left(\dfrac{-it\bar H}
{\hbar}\right)\psi\right]\left(\vec x\right) =
\psi\left(\vec x\right) \text{when } t = 0$, we have
$$\align
{}&\int K_0\left(\vec x, \vec x_0\right)
\phi\left(\vec x\right)\psi\left(\vec x_0\right)
d\vec x_0 d\vec x =
\int_{\Bbb R^n} \phi\left(\vec x\right)
\psi\left(\vec x\right) d\vec x = \tag6.5\\
{}&\iint \phi\left(\vec x\right)
\delta\left(\vec x - \vec x_0\right)
\psi\left(\vec x_0\right) d\vec x\, d\vec x_0.
\endalign
$$
We extend (6.5) to all of $S\left(\Bbb R^n \times \Bbb R^n\right)$.
Let $\eta\left(\vec x, \vec x_0\right)$
$\in S\left(\Bbb R^n \times \Bbb R^n\right)$.
Choose a sequence of functions $\eta{}_k\left(\vec x, \vec x_0\right)$
in $S\left(\Bbb R^n \times \Bbb R^n\right)$
such that $\eta{}_k \to
\eta$ in $S\left(\Bbb R^n \times \Bbb R^n\right)$ and
$\eta{}_k = \sum_{i=0}^{j_k} u_{i,k}
\left(\vec x\right)v_{i,k}\left(\vec x_0\right)$ where
$u_{i,k} \text{ and }v_{i,k} \in D\left(\Bbb R^n\right)$,
the $C^{\infty}$ compactly supported test functions, 
then
$$\align
  {}&\int K_0\left(\vec x, \vec x_0\right)
  \eta\left(\vec x, \vec x_0\right)d\vec x d\vec x_0 =
  \lim_{k\to\infty}
  \int K_0\left(\vec x, \vec x_0\right)
  \eta{}_k\left(\vec x, \vec x_0\right)d\vec x d\vec x_0 
    = \tag6.6\\
  {}&\lim_{k\to\infty}\int_{\Bbb R^n}\sum_{i=0}^{j_k} u_{i,k}
  \left(\vec x\right)v_{i,k}\left(\vec x\right) d\vec x
  =\int_{\Bbb R^n} \eta\left(\vec x, \vec x\right) d\vec x =
  \int \delta\left(\vec x - \vec x_0\right)\eta\left(\vec x, \vec x_0\right)
  d\vec x d\vec x_0 . \qed
\endalign
$$
\enddemo
 
It is well known that the free propagator satisfies
$K_t^{\text{free}}\left(\vec x, \vec x_0\right) =
K_t^{\text{free}}\left(\vec x_0, \vec x\right)$.
We will show a similar property for the tempered distribution
in theorem 6.1.  Intuitively, it is reasonable to believe
that the from lemma 4.5 we can conclude the following.
\proclaim{\bf Theorem 6.4}
$K_t\left(\vec x, \vec x_0\right) =
K_t\left(\vec x_0, \vec x\right)$ where equality
is in the sense of distributions.
\endproclaim
\demo{Proof }$\forall
\phi\left(\vec x\right),
\psi\left(\vec x_0\right) \in S\left(\Bbb R^n\right)
$,
we have that
$$\align
  {}&\int K_t\left(\vec x, \vec x_0\right)
   \phi\left(\vec x\right)\psi\left(\vec x_0\right)d\vec x d\vec x_0 =
   \int_{\Bbb R^n} \phi\left(\vec x\right)\left[
     \text{exp}\left(\dfrac{-it\bar H}{\hbar}\right)
     \psi\right]\left(\vec x\right)
     d\vec x = \tag6.7\\
  {}&\int_{\Bbb R^n} \psi\left(\vec x_0\right)\left[
     \text{exp}\left(\dfrac{-it\bar H}{\hbar}\right)
     \phi\right]\left(\vec x_0\right)
     d\vec x_0 =
    \int_{\Bbb R^n} \psi\left(\vec x\right)\left[
     \text{exp}\left(\dfrac{-it\bar H}{\hbar}\right)
     \phi\right]\left(\vec x\right)
     d\vec x = \\
  {}&\int K_t\left(\vec x, \vec x_0\right)
   \psi\left(\vec x\right)\phi\left(\vec x_0\right)
   d\vec x d\vec x_0 =
   \int K_t\left(\vec x_0, \vec x\right)
   \psi\left(\vec x_0\right)\phi\left(\vec x\right)
   d\vec x d\vec x_0.
\endalign
$$
We extend (6.7) to all of
$S\left(\Bbb R^n \times \Bbb R^n\right)$.
Let $\eta\left(\vec x, \vec x_0\right)
\in S\left(\Bbb R^n \times \Bbb R^n\right)$.
Choose a sequence of functions $\eta{}_k\left(\vec x, \vec x_0\right)$
in $S\left(\Bbb R^{2n} \right)$
such that $\eta{}_k\to
\eta$ in $S\left(\Bbb R^{2n}\right)$
and $\eta{}_k = \sum_{i=0}^{j_k} u_{i,k}
\left(\vec x\right)v_{i,k}\left(\vec x_0\right)$ where
$u_{i,k} \text{ and }v_{i,k} \in D\left(\Bbb R^n\right)$,
the $C^{\infty}$ compactly supported test functions.
We then have 
$$\align
  {}&\int K_t\left(\vec x, \vec x_0\right) 
   \eta\left(\vec x, \vec x_0\right)
   d\vec x d\vec x_0 = \lim_{k\to\infty}
   \int K_t\left(\vec x, \vec x_0\right) 
   \eta{}_k\left(\vec x, \vec x_0\right)
   d\vec x d\vec x_0 = \tag6.8\\
  {}&\lim_{k\to\infty}
  \int K_t\left(\vec x_0, \vec x\right) 
  \eta{}_k\left(\vec x, \vec x_0\right)
  d\vec x d\vec x_0 =
  \int K_t\left(\vec x_0, \vec x\right) 
  \eta\left(\vec x, \vec x_0\right)
  d\vec x d\vec x_0. \qed
\endalign
$$
\enddemo

\subhead{VII Proof of Sixth and Seventh Theorem}
\endsubhead 
\proclaim{Theorem 7.1} The sixth theorem in section II is true.
\endproclaim
\demo{Proof} The nonstandard equivalent of 
equation 1.8 is: for all $\omega\in {}^*\Bbb N - \Bbb N$,
$$
  K\left(\vec x, \vec x_{0},\eta ,\gamma ,t\right) =
  st\left({}^*K_\omega\left(\vec x, \vec x_{0},\eta ,\gamma ,t\right)
   \right). \qed\tag7.1
$$
\enddemo

\proclaim{Theorem 7.2} The seventh theorem in section II is true.
\endproclaim
\demo{Proof} Let
$$\align
  {}&\bar H\left(\eta ,\gamma\right) = 
   \bar{\int}_{\Bbb R^{2n}}\phi\left(\vec x\right)
   \psi\left(\vec x_0\right)
   K\left(\vec x, \vec x_{0},\eta ,\gamma ,t\right)
   \,d\vec x_0 d\vec x, \tag7.2 \\
  {}&H\left(\eta ,\gamma\right) =
   \int_{\Bbb R^{2n}}\phi\left(\vec x\right)
   \psi\left(\vec x_0\right)
   K\left(\vec x, \vec x_{0},\eta ,\gamma ,t\right)
   \,d\vec x_0 d\vec x,\\
  {}&C = \int_{\Bbb R^n} \phi\left(\vec x\right)\left[
  \text{exp}
  \left(\dfrac{-it\bar H}{\hbar}\right)\psi\right]\left(\vec x\right)
  d\vec x. \\
\endalign
$$
The second theorem of section II  
implies that for all $\epsilon \in \Bbb R^{+}$,
there exists a $\delta\in \Bbb R^{+}$ such that
$$\align
  {}&|H\left(\eta ,\gamma\right) - C| < \epsilon \quad when \quad 
  \eta ,\gamma < \delta.  
  \tag7.3\\
\endalign
$$
We now *-transform 7.3 and conclude that any
positive infinitesimal $\eta$ and $\gamma$ is
less than $\delta$.  Since we can do this
for any standard $\epsilon$, the seventh
theorem hold for $H$. A similar argument shows that
the theorem is also true for $\bar H$ \qed
\enddemo

\subhead{VIII The Harmonic Oscillator}
\endsubhead
We now compute the harmonic oscillator propagator for 
$0 < t < \dfrac{\pi}{\lambda}$ using the
formulas above.  Some of the techniques that we will use
was previously worked out in [19], for full details, we will
occassionally refer
the reader to [19].  For the harmonic oscillator, equation 1.5
reads(with a shift in the indices)
$$\align
  {}&K\left(\vec q, \vec q_{0},\eta ,\gamma ,t\right) = \tag8.1\\
  {}&\lim_{k\to\infty}w_{n,k+1}
    \int_{r\Bbb R^{\left(k+2\right)n}}
    F_{\vec q_0}\left(\vec x_0, \gamma\right)
    G_{\vec q}\left(\vec x_{k+1},\eta\right)\times\\
    {}&\text{exp }\left\{\summ{k}{\vec x_j}{\vec x_{j-1}}
       {-\dfrac{m}2\lambda^2 \left(\vec x_j\right)^2}\right\}
       d\vec x_0\dots d\vec x_{k+1}. 
\endalign
$$
Let us write $\vec x_j = \left(x_{j}^1, \dots x_{j}^n\right)$,
and
$$\align
  {}&\text{exp }\left\{\summ{k}{\vec x_j}{\vec x_{j-1}}
       {-\dfrac{m}2\lambda^2 \left(\vec x_j\right)^2}\right\} = \tag8.2\\
  {}&\prod_{\alpha=1}^{n}
    \text{exp }\left\{\summ{k}{x_j^\alpha}{x_{j-1}^\alpha}
       {-\dfrac{m}2\lambda^2 \left(x_j^\alpha\right)^2}\right\}.
\endalign
$$
The popular method to compute the time sliced harmonic oscillator
path integral is to use (8.2) to decouple the integrals in 1.0 
and reduce the problem to produces of one dimensional harmonic
oscillators.   
Due to the extra $\vec x_{0}, \vec x_{k+1}$ integrals in 
8.1, it is not immediately clear that we can use 8.2 to decouple
the improper Riemann integrals.  

For the moment, let us consider just one of the entries in
the product of 8.2.  To
shorten notation, let us write 
$$\align
   {}&\summ{k}{x_j^\alpha}{x_{j-1}^\alpha}
   {-\dfrac{m}2\lambda^2 (x_j^\alpha)^2} = \tag8.3\\
   {}&\left(\dfrac {im}{2\hbar\epsilon}\right)
   \bigg[\left(x_0^\alpha\right)^2 - 2x_0^\alpha x_1^\alpha +
   \left(x_{k+1}^\alpha\right)^2 - 2x_k^\alpha x_{k+1}^\alpha + \\
   {}&\sum\limits_{j = 1}^{k} 2\left(x_j^\alpha\right)^2 -
   \sum\limits_{j = 1}^{k} 2x_j^\alpha x_{j-1}^\alpha -
   \epsilon^2\lambda^2 \sum\limits_{j = 1}^{k+1}
   \left(x_j^\alpha\right)^2\bigg] = \\
\endalign
$$
$$\align
   {}&\left(\dfrac {im}{2\hbar\epsilon}\right)
   \left(\vec x^\alpha\right)^t\Bigg\{
   \left(\!\!\!\!\matrix\format\quad\r&\quad\r&\quad\r&\quad\r&\quad\r\\
   1&-1&0&\hdots&0\\
   -1&{ }&{ }&{ }&\vdots\\
   0&{ }& 0 &{ }&0\\
   \vdots&{ }&{ }&{ }&-1\\
   0&\hdots&{ }&-1&1\\
    \endmatrix\right)  \, -  \, \\
   {}&\epsilon^2\lambda^2
   \left(\matrix
    0&\hdots&{ }&{ }&{ }&\hdots&\hdots&0\\
    0&1&0&\hdots&{ }&{ }&\hdots&0\\
    \vdots&\ddots&\ddots&\ddots&\ddots&{ }&{ }&\vdots\\
    { }&{ }&\ddots&\ddots&\ddots&{ }&{ }&{ }\\
    { }&{ }&{ }&\ddots&\ddots&\ddots&{ }&{ }\\
    \vdots&{ }&{ }&{ }&\ddots&\ddots&{ }&\vdots\\
    0&\hdots&{ }&{ }&\hdots&0&1&0\\
    0&\hdots&{ }&{ }&{ }&\hdots&0&1\\
    \endmatrix\right)\, + \\
   {}&\left(\matrix
   0&\hdots&\hdots&\hdots&\hdots&\hdots&\hdots&\hdots&\hdots&0\\
   0&2&-1&0&\hdots&\hdots&\hdots&\hdots&\hdots&0 \\
   \vdots&-1 & 2 &-1&0&\hdots&\hdots & \hdots&\hdots &0 \\
    \vdots&0&-1&2&-1& 0 &\hdots&\hdots&\hdots&0 \\
    \vdots&\vdots&\ddots&\ddots&\ddots&\ddots&\ddots&\ddots&{ }&\vdots \\
    \vdots&\vdots&{ }&\ddots&\ddots&\ddots&\ddots&\ddots&\ddots&\vdots\\
    0&{ }&{ }&\ddots&\ddots&\ddots&\ddots&\ddots&{ }&{ }\\
    0 & \hdots&\hdots & \hdots &\hdots&0&-1 & 2 & -1 &0\\
    0 & \hdots& \hdots& \hdots & \hdots&\hdots&0&-1&2 &0\\
    0&\hdots&\hdots&\hdots&\hdots&\hdots&\hdots&\hdots&\hdots&0\\
    \endmatrix\right)\Bigg\}\vec x^\alpha =
   \left(\dfrac {im}{2\hbar\epsilon}\right)
   \left(\left(\vec x^\alpha\right)^tT_k\vec x^\alpha\right),
\endalign
$$
where $T_k$ is the $(k+2)$ by $(k+2)$ symmetric
matrix,
$$T_k = \left(\!\!\!\!\matrix\format\quad\r&\quad\r&\quad\r&\quad\r&\quad\r\\
   1&-1&0&\hdots&0\\
  -1&{ }&{ }&{ }&\vdots\\
   0&{ }& S_k &{ }&0\\
   \vdots&{ }&{ }&{ }&-1\\
   0&\hdots&{ }&-1&1 - \epsilon^2\lambda^2\\
    \endmatrix\right),\tag8.4
$$
with $S_k$
being the $k$ by $k$ symmetric matrix
$
   S_k = A_k - \epsilon^2\lambda^2{}B_k
$, where
$$\align
   {}&A_k = \left(\matrix
   2&-1&0&\hdots&\hdots&\hdots&\hdots&0 \\
   -1 & 2 &-1&0&\hdots&\hdots & \hdots &0 \\
    0&-1&2&-1&\hdots&\hdots&\hdots&0 \\
    \vdots&\ddots&\ddots&\ddots&\ddots&\ddots&{ }&\vdots \\
    \vdots&{ }&\ddots&\ddots&\ddots&\ddots&\ddots&\vdots\\
    0&{ }&{ }&\ddots&\ddots&\ddots&\ddots&0\\
    0 & \hdots & \hdots &\hdots&0&-1 & 2 & -1 \\
    0 & \hdots & \hdots & \hdots&\hdots&0&-1&2
    \endmatrix\right), \tag8.5\\
\endalign
$$
$$\align
   {}&B_k = \left(\matrix
    1&0&\hdots&{ }&{ }&{ }&\hdots&0\\
    0&1&0&\hdots&{ }&{ }&\hdots&0\\
    \vdots&\ddots&\ddots&\ddots&{ }&{ }&{ }&\vdots\\
    { }&{ }&\ddots&\ddots&\ddots&{ }&{ }&{ }\\
    { }&{ }&{ }&\ddots&\ddots&\ddots&{ }&{ }\\
    \vdots&{ }&{ }&{ }&\ddots&\ddots&{ }&\vdots\\
    0&\hdots&{ }&{ }&\hdots&0&1&0\\
    0&\hdots&{ }&{ }&{ }&\hdots&0&1\\
\endmatrix\right)\,
\endalign
$$
and
$\vec x^\alpha$ is the column vector
$$
   \vec x^\alpha = \left(\matrix
   x_0^\alpha\\
   x_1^\alpha\\
   \vdots\\
   x_k^\alpha\\
   x_{k+1}^\alpha\\
  \endmatrix\right)\, . \tag8.6
$$
Let $w^\alpha:[0,t]\to\Bbb R$ be such that $|w^\alpha\left(s\right)|<\infty$
and $w^\alpha\left(0\right)=x_0^\alpha, w^\alpha\left(t\right) = x_{k+1}^\alpha$.
In the literature, the path $w^\alpha$ is usually taken to be the 
path of the classical harmonic oscillator.  Here, we allow
$w^\alpha$ to be any finite path that starts at $x_0^\alpha$ and ends
at $x_{k+1}^\alpha$.  We do not assume prior knowledge of classical mechanics.  
We make the substitution $x_j^\alpha =
w^\alpha\left(\frac {jt}{k+1}\right) + y_j^\alpha = w_j^\alpha + y_j^\alpha$
(notice that $y_0^\alpha = 0 = y_{k+1}^\alpha$ 
since $w^\alpha\left(0\right) = x_0^\alpha $
and $w^\alpha\left(t\right) = x_{k+1}^\alpha $).  
Using the fact that $T_k$ is
symmetric, we have
$$\align
   {}&\left(\vec x^\alpha\right)^tT_k\vec x^\alpha = 
    \left(\vec y^\alpha + \vec w^\alpha\right)^tT_k
     \left(\vec y^\alpha + \vec w^\alpha\right) = \tag8.7\\
    {}&\left(\vec w^\alpha\right)^tT_k\vec w^\alpha +
     \left(\vec y^\alpha\right)^tT_k\vec y^\alpha + 
     \left(\vec w^\alpha\right)^tT_k\vec y^\alpha + 
      \left(\vec y^\alpha\right)^tT_k\vec w^\alpha = \\
   {}&\left(\vec w^\alpha\right)^tT_k\vec w^{\alpha} + 
     \left(\vec y^\alpha\right)^tT_k\vec y^\alpha +
     \left(T_k\vec w^\alpha\right)^{t}\vec y^\alpha + 
     \left(\left(\vec w^\alpha\right)^{t}T_k\vec y^\alpha\right)^{t} = \\
    {}&\left(\vec w^\alpha\right)^tT_k\vec w^\alpha + 
    \left(\vec y^\alpha\right)^tT_k\vec y^\alpha +
     \left(T_k\vec w^\alpha\right)^{t}\vec y^\alpha + 
     \left(\vec w^\alpha\right)^{t}T_k\vec y^\alpha = \\
   {}&\left(\vec w^\alpha\right)^tT_k\vec w^\alpha + 
    \left(\vec y^\alpha\right)^tT_k\vec y^\alpha +
     2\left(T_k\vec w^\alpha\right)^t\vec y^\alpha,
\endalign
$$
where
$$
   \vec y^\alpha = \left(\matrix
     0\\
    y_1^\alpha\\
    \vdots\\
    y_k^\alpha\\
      0\\
   \endmatrix\right), \quad
   \vec w^\alpha = \left(\matrix
     w_0^\alpha = x_0^\alpha\\
    w_1^\alpha\\
    \vdots\\
    w_k^\alpha\\
      w_{k+1}^\alpha = x_{k+1}^\alpha\\
   \endmatrix\right).
   \tag8.8
$$
By using $y_0^\alpha = 0 = y_{k+1}^\alpha$ and writing $T_k$ as
$$ T_k =
   \left(\!\!\!\!\matrix\format\quad\r&\quad\r&\quad\r&\quad\r&\quad\r\\
   0&0&0&\hdots&0\\
   0&{ }&{ }&{ }&\vdots\\
   0&{ }& S_k &{ }&0\\
   \vdots&{ }&{ }&{ }&0\\
   0&\hdots&{ }&0&0\\
    \endmatrix\right) +
   \left(\!\!\!\!\matrix\format\quad\r&\quad\r&\quad\r&\quad\r&\quad\r\\
   1&-1&0&\hdots&0\\
  -1&{ }&{ }&{ }&\vdots\\
   0&{ }& 0 &{ }&0\\
   \vdots&{ }&{ }&{ }&-1\\
   0&\hdots&{ }&-1&1 - \epsilon^2\lambda^2\\
    \endmatrix\right),\tag8.9
$$
we obtain
$$\align
   {}&\left(\vec x^\alpha\right)^tT_k\vec x^\alpha = 
   \left(\vec w^\alpha\right)^tT_k\vec w^\alpha + 
   \left(\vec y^\alpha\right)^tT_k\vec y^\alpha +
     2\left(T_k\vec w^\alpha\right)^t\vec y^\alpha = \tag8.10\\
   {}&\left(\vec w^\alpha\right)^tT_k\vec w^\alpha + 
    \left(\hat y^\alpha\right)^tS_k\hat y^\alpha +
     2\left(\vec \rho^\alpha\right)^t\hat y^\alpha\, 
\endalign
$$ where
$$\align
   {}&\hat y^\alpha = \left(\matrix
    y_1^\alpha\\
    \vdots\\
     y_k^\alpha
   \endmatrix\right), \quad
   \vec \rho^\alpha = S_k\left(\matrix
     w_1^\alpha\\
      \vdots\\
       w_{k-1}^\alpha\\
       w_k^\alpha\\
   \endmatrix\right) - \left(\matrix
       w_0^\alpha = x_0^\alpha\\
       0\\
       \vdots\\
       0\\
       w_{k+1}^\alpha = x_{k+1}^\alpha \\
   \endmatrix\right)\, = S_k\hat w^\alpha - \hat{\hat w}^{\alpha} .
   \tag8.11
\endalign
$$
\proclaim{\bf Lemma 8.1} Let $t\in\Bbb R$ and
$0 < t < \dfrac{\pi}{\lambda}$.  For any
$\omega\in{}^*\Bbb N - \Bbb N, {}^{*}S_{\omega}$ is
positive definite in the $*$-transformed sense.  Here,
${}^{*}S_{\omega}$ is the *-transform of the matrix
$S_k$ defined after equation 8.4.
\endproclaim
\demo{Proof} See [19].
\enddemo
We now go back to equations 8.1 and 8.2 in Nonstandard Analysis
form.  With an abuse of notation, in Nonstandard Analysis 8.1 reads
$$\align
  {}&K\left(\vec q, \vec q_{0},\eta ,\gamma ,t\right) = \tag8.12\\
  {}&st\Bigg\{w_{n,\omega + 1}
    \int_{r\Bbb R^{\left(\omega+2\right)n}}
    F_{\vec q_0}\left(\vec x_0, \gamma\right)
    G_{\vec q}\left(\vec x_{\omega+1},\eta\right)\times\\
    {}&\text{exp }\left\{\summ{\omega}{\vec x_j}{\vec x_{j-1}}
       {-\dfrac{m}2\lambda^2 \left(\vec x_j\right)^2}\right\}
       d\vec x_0\dots d\vec x_{\omega+1}\Bigg\} = \\
    {}&st\Bigg\{w_{n,\omega + 1}
    \int_{r\Bbb R^{\left(\omega+2\right)n}}
    F_{\vec q_0}\left(\vec x_0, \gamma\right)
    G_{\vec q}\left(\vec x_{\omega+1},\eta\right)\times\\
    {}&\prod_{\alpha=1}^{n}
    \text{exp }\left\{\summ{\omega}{x_j^\alpha}{x_{j-1}^\alpha}
       {-\dfrac{m}2\lambda^2 \left(x_j^\alpha\right)^2}\right\}
       d\vec x_0\dots d\vec x_{\omega+1}\Bigg\} = \\
    {}&st\Bigg\{w_{n,\omega + 1}
    \int_{r\Bbb R^{\left(\omega+2\right)n}}
    F_{\vec q_0}\left(\vec x_0, \gamma\right)
    G_{\vec q}\left(\vec x_{\omega+1},\eta\right)\times\\
    {}&\prod_{\alpha=1}^{n}
    \text{exp }\left\{\left(\dfrac {im}{2\hbar\epsilon}\right)
   \left(\left(\vec x^\alpha\right)^tT_\omega\vec x^\alpha\right)
    \right\}
       d\vec x_0\dots d\vec x_{\omega+1}\Bigg\}. 
\endalign
$$
We perform a *-transform of the change of variable 
described in equation 8.7 on
$\vec x_1\dots\vec x_\omega$, and obtain
$$\align
  {}&K\left(\vec q, \vec q_{0},\eta ,\gamma ,t\right) = \tag8.13\\
   {}&st\Bigg\{w_{n,\omega + 1}
    \int_{r\Bbb R^{\left(\omega+2\right)n}}
    F_{\vec q_0}\left(\vec x_0, \gamma\right)
    G_{\vec q}\left(\vec x_{\omega+1},\eta\right)\times\\
    {}&\prod_{\alpha=1}^{n}
    \text{exp }\left\{\left(\dfrac {im}{2\hbar\epsilon}\right)
    \left(\vec w^\alpha\right)^tT_\omega\vec w^\alpha\right\} 
     \times\\
    {}&\prod_{\alpha=1}^{n}
    \text{exp }\left\{\left(\dfrac {im}{2\hbar\epsilon}\right)
    \left[\left(\hat y^\alpha\right)^tS_\omega\hat y^\alpha +
     2\left(\vec \rho^\alpha\right)^t\hat y^\alpha\right]
    \right\}
       d\vec x_0d\vec x_{\omega+1}d\vec y_1\dots d\vec y_{\omega}\Bigg\}.
\endalign
$$
Since $S_\omega$ is positive definite, it is invertable.
Since $S_\omega$ is symmetric, the following is true
$$
  \left(\hat y^\alpha\right)^tS_\omega\hat y^\alpha +
   2\left(\vec \rho^\alpha\right)^t\hat y^\alpha
   = \left(\hat y^\alpha + S_\omega^{-1}\vec \rho^\alpha\right)^tS_\omega
  \left(\hat y^\alpha + S_\omega^{-1}\vec \rho^\alpha\right) -
   \left(\vec \rho^\alpha\right)^tS_\omega^{-1}\vec \rho^\alpha. 
\tag8.14$$
Using 8.14 in 8.13 and performing the 
transformation $z_j^\alpha = y_j^\alpha +
\left(S_\omega^{-1}\vec \rho^\alpha\right)_j$,
we obtain
$$\align
  {}&K\left(\vec q, \vec q_{0},\eta ,\gamma ,t\right) = \tag8.15\\
   {}&st\Bigg\{w_{n,\omega + 1}
      \int_{r\Bbb R^{\left(\omega+2\right)n}}
    F_{\vec q_0}\left(\vec x_0, \gamma\right)
    G_{\vec q}\left(\vec x_{\omega+1},\eta\right)\times\\
    {}&\prod_{\alpha=1}^{n}
    \text{exp }\left\{\left(\dfrac {im}{2\hbar\epsilon}\right)
    \left[\left(\vec w^\alpha\right)^tT_\omega\vec w^\alpha - 
   \left(\vec \rho^\alpha\right)^tS_\omega^{-1}\vec \rho^\alpha
    \right]\right\}
     \times\\
    {}&\prod_{\alpha=1}^{n}
    \text{exp }\left\{\left(\dfrac {im}{2\hbar\epsilon}\right)
    \left(\vec z^\alpha\right)^tS_\omega\vec z^\alpha 
    \right\}
       d\vec x_0d\vec x_{\omega+1}d\vec z_1\dots d\vec z_{\omega}\Bigg\}.
\endalign
$$
Notice that before the improper limits are taken on the integrals,
the limits of integration on the variables $z_j^\alpha$ are dependent
on $\vec x_0$ and $\vec x_{\omega + 1}$ due to the fact that
$\vec \rho^\alpha$ is dependent on them.  Thus, we still can not
decouple the improper Riemann integrals.

Let us take a look at the limits of integrations more
closely.
Before the improper limits on the integrals are taken in 
equation 8.15, we have
$$\align
   {}&w_{n,\omega + 1}\int_{\Cal O}
    F_{\vec q_0}\left(\vec x_0, \gamma\right)
    G_{\vec q}\left(\vec x_{\omega+1},\eta\right)
    \times\tag8.16\\
   {}&\prod_{\alpha=1}^{n}
    \text{exp }\left\{\left(\dfrac {im}{2\hbar\epsilon}\right)
    \left[\left(\vec w^\alpha\right)^tT_\omega\vec w^\alpha -
   \left(\vec \rho^\alpha\right)^tS_\omega^{-1}\vec \rho^\alpha
      \right]\right\}
     \times\\
   {}&\Bigg\{\int_{\Cal {\bar O} }
    \prod_{\alpha=1}^{n}
    \text{exp }\left\{\left(\dfrac {im}{2\hbar\epsilon}\right)
    \left(\vec z^\alpha\right)^tS_\omega\vec z^\alpha
    \right\}
       d\vec z_1\dots d\vec z_{\omega}\Bigg\} 
       d\vec x_0d\vec x_{\omega+1} ,
\endalign
$$
where both $\Cal O$ and $\Cal{\bar O}$
are *-compact and the boundry of
$\Cal{\bar O}$ depends on $\vec x_0$ 
, $\vec x_{\omega + 1}$ and a set of 
indices $\left\{J\right\}$ such that
as $\left\{J\right\}\to\infty$,
$\Cal{\bar O} \to {}^*\Bbb R^{\omega n}$
in the *-transformed sense.
The reason for $\Cal{\bar O}$'s dependent on
$\vec x_0$
, $\vec x_{\omega + 1}$ is due to the fact
that $\vec \rho^\alpha$ is dependent on them
(equation 8.11)
and we performed the change of variables from
equation 8.13 to equation 8.15.  Further, 
the boundary of $\Cal O$ is also indexed by
a set similar to that of $\left\{J\right\}$.
What we would
like to do is pass the $\left\{J\right\}$
limits inside the $\Cal{O}$ integral and
decouple the improper Riemann integrals in
equation 8.15 into improper Riemann integrals in
$d\vec z_1\dots d\vec z_\omega$ then 
$d\vec x_0 d\vec x_{\omega + 1}$.

It is well known from the one dimensional harmonic
oscillator that 
$$\align
   {}&w_{1,k+1}\int\limits_{r\Bbb R^k}
     \text{exp}\left(\dfrac{im}{2\hbar\epsilon}
     \left(z^\alpha\right)^tS_kz^\alpha\right)
       dz_1^\alpha\dots dz_k^\alpha = 
      \left(\dfrac{m}{2\pi i\hbar\epsilon}\right)^{\frac{1}{2}}
       \sqrt{\dfrac{1}{\text{ det } S_k}}.
     \tag8.17
\endalign
$$
Let us fix an $\Cal O$.
Since $\Cal O$ is compact
(see the construction in equation 3.12). 
We can *-transform and
conclude from 8.16
and 8.17 that for any $\beta \in {}^*\Bbb R^{+}$, 
there exists a fixed
$M \in {}^*\Bbb R^{+}$
that depends only on $\Cal O$ such that
$$\align
   {}&
   \Bigg |w_{n,\omega + 1}
   \int_{\Cal {\bar O} }
    \prod_{\alpha=1}^{n}
    \text{exp }\left\{\left(\dfrac {im}{2\hbar\epsilon}\right)
    \left(\vec z^\alpha\right)^tS_\omega\vec z^\alpha
    \right\}
       d\vec z_1\dots d\vec z_{\omega} - \tag8.18\\
   {}&\left( \left(\dfrac{m}{2\pi i\hbar\epsilon}\right)^{\frac{1}{2}}
       \sqrt{\dfrac{1}{\text{ det } S_\omega}} 
    \right)^n \Bigg |< \beta
\endalign
$$
whenever all entries of $\left\{J\right\}$ are
bigger than $M$.  In other words, because
$\Cal O$ is compact, for all
$\left(\vec x_0,\vec x_{\omega +1}\right) \in \Cal O$,
equation 8.18 is true whenever all entries of 
$\left\{J\right\}$ are
bigger than a fixed $M$;
further, this $M$ depends on
$\Cal O$. 

Equation 8.18 allows us to use *-Lebesgue dominating
convergence theorem and pass the $\left\{J\right\}$
limits inside the $\Cal O$ integral and decouples the improper 
Riemann integrals.  Thus, we have proved the following
\proclaim{Theorem 8.2} For the harmonic oscillator,
$$\align
  {}&K\left(\vec q, \vec q_{0},\eta ,\gamma ,t\right) = \tag8.19\\
   {}&st\Bigg\{w_{n,\omega + 1}
      \int_{r\Bbb R^{\left(\omega+2\right)n}}
    F_{\vec q_0}\left(\vec x_0, \gamma\right)
    G_{\vec q}\left(\vec x_{\omega+1},\eta\right)\times\\
    {}&\prod_{\alpha=1}^{n}
    \text{exp }\left\{\left(\dfrac {im}{2\hbar\epsilon}\right)
    \left[\left(\vec w^\alpha\right)^tT_\omega\vec w^\alpha -
   \left(\vec \rho^\alpha\right)^tS_\omega^{-1}\vec \rho^\alpha
    \right]\right\}
     \times\\
 {}&\prod_{\alpha=1}^{n}
    \text{exp }\left\{\left(\dfrac {im}{2\hbar\epsilon}\right)
    \left(\vec z^\alpha\right)^tS_\omega\vec z^\alpha
    \right\}
       d\vec x_0d\vec x_{\omega+1}d\vec z_1
    \dots d\vec z_{\omega}\Bigg\} = \\
 {}&st\Bigg\{w_{n,\omega + 1}
      \Bigg\{\int_{r\Bbb R^{2n} }
    F_{\vec q_0}\left(\vec x_0, \gamma\right)
    G_{\vec q}\left(\vec x_{\omega+1},\eta\right)\times\\
    {}&\prod_{\alpha=1}^{n}
    \text{exp }\left\{\left(\dfrac {im}{2\hbar\epsilon}\right)
    \left[\left(\vec w^\alpha\right)^tT_\omega\vec w^\alpha -
   \left(\vec \rho^\alpha\right)^tS_\omega^{-1}\vec \rho^\alpha
    \right]\right\}
    d\vec x_0\vec x_{\omega + 1}
     \Bigg\}\times\\
   {}&\int_{r\Bbb R^{\omega n}}
   \prod_{\alpha=1}^{n}
    \text{exp }\left\{\left(\dfrac {im}{2\hbar\epsilon}\right)
    \left(\vec z^\alpha\right)^tS_\omega\vec z^\alpha
    \right\}
    d\vec z_1
    \dots d\vec z_{\omega}\Bigg\} = \\
    {}&st\Bigg\{
       \left( \left(\dfrac{m}{2\pi i\hbar\epsilon}\right)^{\frac{1}{2}}
       \sqrt{\dfrac{1}{\text{ det } S_\omega}}
       \right)^n
        \int_{r\Bbb R^{2n} }
    F_{\vec q_0}\left(\vec x_0, \gamma\right)
    G_{\vec q}\left(\vec x_{\omega+1},\eta\right)\times\\
    {}&\prod_{\alpha=1}^{n}
    \text{exp }\left\{\left(\dfrac {im}{2\hbar\epsilon}\right)
    \left[\left(\vec w^\alpha\right)^tT_\omega\vec w^\alpha -
   \left(\vec \rho^\alpha\right)^tS_\omega^{-1}\vec \rho^\alpha
    \right]\right\}
    d\vec x_0\vec x_{\omega + 1}
   \Bigg\}.
\endalign
$$
\endproclaim
\demo{Proof} See above.  \qed
\enddemo
It remains now to compute the last equality in 8.19.
\proclaim{Proposition 8.3}With the previously defined notations,
$$\align
  {}&st\left\{\left( \left(\dfrac{m}{2\pi i\hbar\epsilon}\right)^{\frac{1}{2}}
       \sqrt{\dfrac{1}{\text{ det } S_\omega}}
       \right)^n\right\} = 
        \left(\dfrac{m}{2\pi{}i\hbar}\right)^{\frac{n}2}
        \left(\dfrac{\lambda}
        {\sin\lambda{}t}\right)^{\frac{n}{2}}.
  \tag8.20
\endalign
$$
\endproclaim
\demo{Proof} See [19].  \qed
\enddemo
\proclaim{Proposition 8.4} Let 
$\vec x_0, \vec x_n = \vec y\in \Bbb R^n$ be fixed.
With the previously defined
notations,
$$\align
  {}&\lim_{k\to\infty}\left\{\prod_{\alpha=1}^{n}
    \text{exp }\left\{\left(\dfrac {im}{2\hbar\epsilon}\right)
    \left[\left(\vec w^\alpha\right)^tT_k\vec w^\alpha -
   \left(\vec \rho^\alpha\right)^tS_k^{-1}\vec \rho^\alpha
   \right]\right\}
   \right\} = \tag8.21 \\
   {}&\text{exp}\left\{\dfrac{im}{\hbar}\dfrac{\lambda}{\sin\lambda{}t}
\left[\left(\vec x_0^2 + \vec y^2\right)\cos\lambda{}t - 
  2\vec y\vec x_0\right]
\right\}. 
\endalign
$$
\endproclaim
\remark{Remark 8.5}
Notice that proposition 8.4 does not take place in the nonstandard
world.
\endremark
\demo{Proof} This is just the classical version of the
nonstandard results obtained from [19].  See [19] for
more details.
\enddemo

\proclaim{Proposition 8.5} With the previously defined
notations,
$$\align
   {}&st\Bigg\{\int_{r\Bbb R^{2n} }
    F_{\vec q_0}\left(\vec x_0, \gamma\right)
    G_{\vec q}\left(\vec x_{\omega+1},\eta\right)\times\tag8.22\\
    {}&\prod_{\alpha=1}^{n}
    \text{exp }\left\{\left(\dfrac {im}{2\hbar\epsilon}\right)
    \left(\vec w^\alpha\right)^tT_\omega\vec w^\alpha -
   \left(\vec \rho^\alpha\right)^tS_\omega^{-1}\vec \rho^\alpha\right\}
    d\vec x_0\vec x_{\omega + 1}\Bigg\} = \\
  {}&\int_{\Bbb R^{2n} }
    F_{\vec q_0}\left(\vec x_0, \gamma\right)
    G_{\vec q}\left(\vec y,\eta\right)\times\\
  {}&\text{exp}\left\{\dfrac{im}{\hbar}\dfrac{\lambda}{\sin\lambda{}t}
    \left[\left(\vec x_0^2 + \vec y^2\right)\cos\lambda{}t -
    2\vec y\vec x_0\right]
    \right\}d\vec x_0 d\vec y. 
\endalign
$$
\endproclaim
\demo{Proof} Using proposition 8.4 and Lebesgue's dominating
convergence theorem, we obtain
$$\align
   {}&st\Bigg\{\int_{r\Bbb R^{2n} }
    F_{\vec q_0}\left(\vec x_0, \gamma\right)
    G_{\vec q}\left(\vec x_{\omega+1},\eta\right)\times\tag8.23\\
    {}&\prod_{\alpha=1}^{n}
    \text{exp }\left\{\left(\dfrac {im}{2\hbar\epsilon}\right)
    \left[\left(\vec w^\alpha\right)^tT_\omega\vec w^\alpha -
   \left(\vec \rho^\alpha\right)^tS_\omega^{-1}\vec \rho^\alpha
    \right]\right\} 
    d\vec x_0\vec x_{\omega + 1}\Bigg\} = \\
   {}&\lim_{k\to\infty}
    \int_{\Bbb R^{2n} }
    F_{\vec q_0}\left(\vec x_0, \gamma\right)
    G_{\vec q}\left(\vec y,\eta\right)\times \\
    {}&\prod_{\alpha=1}^{n}
    \text{exp }\left\{\left(\dfrac {im}{2\hbar\epsilon}\right)
    \left[\left(\vec w^\alpha\right)^tT_k\vec w^\alpha -
   \left(\vec \rho^\alpha\right)^tS_k^{-1}\vec \rho^\alpha
   \right]\right\}
   d\vec x_0 d\vec y = \\
  {}&\int_{\Bbb R^{2n} }
    F_{\vec q_0}\left(\vec x_0, \gamma\right)
    G_{\vec q}\left(\vec y,\eta\right)\times \\
    {}&\lim_{k\to\infty}
    \left\{\prod_{\alpha=1}^{n}
    \text{exp }\left\{\left(\dfrac {im}{2\hbar\epsilon}\right)
    \left[\left(\vec w^\alpha\right)^tT_k\vec w^\alpha -
   \left(\vec \rho^\alpha\right)^tS_k^{-1}\vec \rho^\alpha
    \right]\right\}
   \right\}d\vec x_0 d\vec y = \\
    {}&\int_{\Bbb R^{2n} }
    F_{\vec q_0}\left(\vec x_0, \gamma\right)
    G_{\vec q}\left(\vec y,\eta\right) 
   \text{exp}\left\{\dfrac{im}{\hbar}\dfrac{\lambda}{\sin\lambda{}t}
\left[\left(\vec x_0^2 + \vec y^2\right)\cos\lambda{}t -
  2\vec y\vec x_0\right]
\right\}d\vec x_0 d\vec y. \qed
\endalign
$$
\enddemo
\proclaim{Theorem 8.6} For the harmonic oscillator,
$$\align
  {}&K\left(\vec q, \vec q_{0},\eta ,\gamma ,t\right) = \tag8.24\\
  {}&\left(\dfrac{m}{2\pi{}i\hbar}\right)^{\frac{n}2}
        \left(\dfrac{\lambda}
        {\sin\lambda{}t}\right)^{\frac{n}{2}}\times \\
  {}&\int_{\Bbb R^{2n} }
    F_{\vec q_0}\left(\vec x_0, \gamma\right)
    G_{\vec q}\left(\vec y,\eta\right)
   \text{exp}\left\{\dfrac{im}{\hbar}\dfrac{\lambda}{\sin\lambda{}t}
\left[\left(\vec x_0^2 + \vec y^2\right)\cos\lambda{}t -
  2\vec y\vec x_0\right]
\right\}d\vec x_0 d\vec y.
\endalign
$$
\endproclaim
\demo{Proof} Equation 8.24 follows from theorem 8.2,
  proposition 8.3, and proposition 8.5. \qed
\enddemo

\proclaim{Theorem 8.7} Let $\phi ,\psi\in L^1\cap L^2$, 
then for the harmonic oscillator Hamiltonian and for
$0 < t < \pi/\lambda$,
$$\align
  {}&\int_{\Bbb R^n} \phi\left(\vec x\right)\left[
  \text{exp}
  \left(\dfrac{-it\bar H}{\hbar}\right)\psi\right]\left(\vec x\right)
  d\vec x  = \tag8.25 \\
  {}&\left(\dfrac{m}{2\pi{}i\hbar}\right)^{\frac{n}2}
        \left(\dfrac{\lambda}
        {\sin\lambda{}t}\right)^{\frac{n}{2}}\times\\
 {}&\int_{\Bbb R^{2n}} \phi\left(\vec q_0\right)
  \psi\left(\vec q\right)
   \text{exp}\left\{\dfrac{im}{\hbar}\dfrac{\lambda}{\sin\lambda{}t}
\left[\left(\vec q_0^2 + \vec q^2\right)\cos\lambda{}t -
  2\vec q\vec q_0\right]
\right\}d\vec q_0 d\vec q.
\endalign
$$
\endproclaim
\demo{Proof} Notice that 
$|K\left(\vec q, \vec q_{0},\eta ,\gamma ,t\right)| 
\leq C_{n,\lambda ,t}$.
Substituting equation 8.24 in equation 1.3 and using Lebesgue
dominating theorem on the $\eta ,\gamma$ limits give 8.25. \qed
\enddemo

\enddocument